\documentclass[aps,showpacs,twocolumn,preprintnumbers,multicol]{revtex4}
\usepackage{epsfig}
\usepackage{color}
\usepackage{amssymb,amsmath}

\def\cA{{\cal A}}
\def\cB{{\cal B}}

\def\cF{{\cal F}}

\def\cJ{{\cal J}}

\def\tr{\mathop{\rm tr}}

\newcommand{\beq}{\begin{equation}}
\newcommand{\eeq}{\end{equation}}

\def\half{{1\over2}}

\newcommand{\tmfloatcontents}{}
\newlength{\tmfloatwidth}
\newcommand{\tmfloat}[5]{
  \renewcommand{\tmfloatcontents}{#4}
  \setlength{\tmfloatwidth}{\widthof{\tmfloatcontents}+1in}
  \ifthenelse{\equal{#2}{small}}
    {\ifthenelse{\lengthtest{\tmfloatwidth > \linewidth}}
      {\setlength{\tmfloatwidth}{\linewidth}}{}}
    {\setlength{\tmfloatwidth}{\linewidth}}  \begin{minipage}[#1]{\tmfloatwidth}
    \begin{center}
      \tmfloatcontents
      \captionof{#3}{#5}
    \end{center}
  \end{minipage}}

\begin{document}

\newpage
\title{ The hadronic light by light contribution to the $(g-2)_{\mu}$ with holographic models of QCD}

\author{ Luigi~Cappiello$^{1,2}$, Oscar~Cat\`a$^3$ and Giancarlo~D'Ambrosio$^2$}

\affiliation{$^1$Dipartimento di Scienze Fisiche,
             Universit\'a di Napoli "Federico II", Via Cintia, 80126 Napoli, Italy}
\affiliation{$^2$INFN-Sezione di Napoli, Via Cintia, 80126 Napoli, Italy}
\affiliation{$^3$Departament de F\'isica Te\`orica and IFIC, Universitat de Val\`encia-CSIC,
Apt. Correus 22085, E-46071 Val\`encia, Spain}

\begin{abstract}
We study the anomalous electromagnetic pion form factor
$F_{\pi^0\gamma^*\gamma^*}$ with a set of holographic models. By comparing with the measured value of the
linear slope, some of these models can be ruled out. From the remaining models we obtain
 predictions  for the low-energy quadratic slope parameters of
$F_{\pi^0\gamma^*\gamma^*}$, currently out of experimental reach
but testable in the near future. We find it particularly useful to
encode this low-energy information in a form factor able to
satisfy also QCD short-distance constraints. We choose the form factor
introduced by D'Ambrosio, Isidori and Portoles in kaon decays, which has the right short distance for a particular value of the quadratic slope, which is later shown to be compatible with our holographic predictions. We then turn to a
determination of the (dominant) pion exchange diagram in the
hadronic light by light scattering contribution to the muon
anomalous magnetic moment. We quantify the theoretical uncertainty
in $(g-2)_{\mu}$ coming from the different input we use: QCD short
distances, experimental input and low-energy holographic
predictions. We also test the pion-pole approximation. Our final
result is $a_{\mu}^{\pi^0}=6.54(25)\cdot 10^{-10}$, where the
error is driven by the linear slope of $F_{\pi^0\gamma^*\gamma^*}$, soon
to be measured with precision at KLOE-2. Our numerical analysis
also indicates that large values of the magnetic susceptibility
$\chi_0$ are disfavored, therefore pointing at a mild effect from
the pion off-shellness. However, in the absence of stronger bounds
on $\chi_0$, an additional $(10-15)\%$ systematic uncertainty on
the previous value for $a_\mu^{\pi^0}$ cannot be excluded.
\end{abstract}
\maketitle
\section{Introduction}\label{secI}
The anomalous magnetic moment of the muon is presently one of the most stringent tests of
the standard model and, due to the current experimental precision, a very useful tool in the
search for new physics. As of today, a discrepancy of roughly three standard deviations
persists between the standard model estimate and the experimental value at
Brookhaven~\cite{Bennett:2004pv}. Whether this is to be ascribed to evidence for
new physics depends crucially on the reliability of the standard model computation.

With the electroweak corrections under good control, the main source of uncertainty on the
theoretical side comes from the hadronic contributions, namely the vacuum and the
light-by-light contributions. While information on the hadronic vacuum polarization
can be extracted from experimental data on the hadronic $e^+e^-$ cross section, the
hadronic light-by-light (HLBL) contribution can only be estimated through nonperturbative
theoretical methods, {\it{i.e.}} lattice simulations and hadronic models.
Due to the complexity of the calculation, a lattice result is unfortunately still
not available and one has to resort to the latter.

In the past there have been plenty of determinations based on
particular models like the ENJL~\cite{Bijnens:2001cq},
HLS~\cite{Hayakawa:1997rq} and more recently on different models
inspired by the $1/N_c$ expansion under the assumption of lowest
meson dominance
(LMD)~\cite{Knecht:2001qf,Melnikov:2003xd,Nyffeler:2009tw}.
Different as they are in nature, all the abovementioned models
eventually determine their free parameters imposing constraints,
either theoretical or experimental. Therefore, the source of
uncertainty generally depends on: (i) the experimental input used;
(ii) the different theoretical constraints imposed; and (iii) an
intrinsic model dependence, which is very difficult to estimate.
While it is certainly reassuring that, ever since the seminal
paper of Ref.~\cite{Knecht:2001qg} settled an important sign
issue, the results for the different models lie in the same
ballpark and no big disagreement is present, there are certain
unresolved issues that might affect the theoretical error estimate
(see for instance the recent discussion
in~\cite{Nyffeler:2010rd}). With the proposals at
Fermilab~\cite{Carey:2009zzb}, J-PARC~\cite{Imazato:2004fy} and
potentially at Frascati~\cite{Babusci:2010ym}, the increase in
experimental precision will aim at reaching $1.5\cdot 10^{-10}$
accuracy. Therefore, a reliable estimate of the theoretical error
in the hadronic light-by-light contribution becomes of the utmost
importance (see, for instance, the recent assessments of the error
budget given in Refs.~\cite{Prades:2009tw,Jegerlehner:2009ry}).

In this paper we propose to use the holographic
principle~\cite{Maldacena:1997re} to address such issues.
Holographic models of QCD, while also based on the $1/N_c$
expansion, offer several advantages over four-dimensional LMD
models. First and foremost, AdS/QCD models are implemented at the
Lagrangian level, and therefore Ward identities between
correlators are automatically fulfilled. Second, due to the AdS
metric, short-distance matching to asymptotically free QCD is
easily achieved. Third, hadronic resonances in AdS/QCD models
arise as the Kaluza-Klein states in the process of
compactification from five to four dimensions. Therefore, as
opposed to LMD models, the full tower of states is automatically
implemented and its separate contributions can be analyzed.
Finally, even though an infinite number of resonances is present,
the number of free parameters is very small. Thus, unlike
phenomenological hadronic models, very little input is needed in
order to be predictive. Holography, therefore, has the potential
to become a consistent hadronic model for all the hadronic
light-by-light contributions to $(g-2)_{\mu}$, at least at leading
order in the $1/N_c$ expansion, {\it{i.e.}}, in the limit of
single-resonance exchange.

In this paper we will concentrate on the neutral pion exchange
contribution. Traditionally, this contribution has been extracted
from the knowledge on the electromagnetic pion form factor
$F_{\pi^0\gamma^*\gamma^*}$. The questions we want to address are the following:

\begin{itemize}
\item [(i)] Which are the parameters entering
$F_{\pi^0\gamma^*\gamma^*}$ that mostly affect the uncertainty on
$(g-2)_{\mu}$?

\item [(ii)] It has been recently pointed out~\cite{Gino09} that the quadratic slope $\hat\beta$, defined by
\begin{eqnarray}\label{lowe}
\lim_{Q_1^2,Q_2^2\to 0}F_{\pi^0\gamma^*\gamma^*}(Q_{1}^{2},Q_{2}^{2})\simeq -\frac{N_C}{12\pi^2f_{\pi}}\times\,\,\,\,\,\,\,\,\,\,\,\,\,\,\,\,\,\,\,\,\,\,\,\,\,&&\nonumber\\
\left[1+{\hat{\alpha}}~(Q_1^2+Q_2^2)+{\hat{\beta}}~Q_1^2Q_2^2+{\hat{\gamma}}~(Q_1^4+Q_2^4)+\cdots\right]&&\nonumber
\end{eqnarray}
might be of importance to reduce the uncertainty in the $(g-2)_\mu$. However, an experimental determination thereof is still lacking. Can holographic models of QCD be used to estimate the linear and quadratic slopes of
$F_{\pi^0\gamma^*\gamma^*}$? Do these models satisfy the required short-distance constraints? How big is the impact of the holographic slopes on $(g-2)_{\mu}$?

\item [(iii)]  Is vector-meson dominance a justified approximation for $F_{\pi^0\gamma^*\gamma^*}$? How does this approximation carry over to the evaluation of the $(g-2)_{\mu}$? 
 
\item [(iv)] What is the impact of not restricting the neutral pion to be on-shell in the HLBL contribution? In other words, what is the accuracy of the pion-pole approximation?
\end{itemize}

In order to address these questions we will study $F_{\pi^0\gamma^*\gamma^*}$ in a set of holographic models, which mostly differ in the way they implement chiral symmetry breaking and the spacing in the hadronic vector spectrum (Regge-like or not). We will test the models against existing experimental data on $F_{\pi^0\gamma^*\gamma^*}$.
In particular, we will compare the different predictions for the linear
slope $\hat\alpha$ with the results from CELLO~\cite{Behrend:1990sr}.
Agreement with experiment will be used as a filtering criteria for
the different models. Then, from the accepted models, we will
extract a prediction for the quadratic slopes $\hat\beta$ and $\hat\gamma$.

For the analysis of the $(g-2)_{\mu}$ we will adopt a strategy similar in spirit to Ref.~\cite{Bijnens:1999jp}, where a set of simple interpolators were
tested against experiment and then used to estimate the
contribution to the HLBL. A particularly useful form factor, which
displays immediately the required short and long-distance
properties is found to be the one used by D'Ambrosio, Isidori and
Portoles (DIP) in kaon decays \cite{D'Ambrosio:1997jp}: 
\begin{eqnarray}\label{intpolr}
K(q_1^2,q_2^2)&=&1+\lambda\left(\frac{q_1^2}{q_1^2-m_{V}^2}+\frac{q_2^2}{q_2^2-m_{V}^2}\right)\nonumber\\
&+&\eta\frac{q_1^2 q_2^2}{(q_1^2-m_{V}^2)(q_2^2-m_{V}^2)}~.\nonumber
\end{eqnarray}
In particular, to study the pion off-shellness we will promote the DIP form factor to an interpolator valid for arbitrary pion momentum. This leads to a new short-distance constraint~\cite{Nyffeler:2009tw} which is naturally implemented in our DIP interpolator.
  
This paper is organized as follows: in Section~\ref{secII}  we
will briefly review the holographic principle and its most common
realizations to study QCD. These models will then be used in
Section~\ref{secIII} to study the $\pi^0\gamma^*\gamma^*$ form factor,
with special emphasis on its low-energy predictions. In
Section~\ref{secIV} we will introduce the DIP interpolator and i) give a first estimate of the pion
exchange contribution to the HLBL piece of $(g-2)_{\mu}$ and then ii)
present an extension of the DIP interpolator with an extra pole,
which will allow us to play with the whole set of long and short-distance constraints and test the stability of our results. Finally, conclusions will be given in
Section~\ref{secV}. Technical details are provided in two
Appendices. The paper is organized such that readers interested in phenomenological applications can skip Section~\ref{secII} without loss of continuity.


\section{Holographic models of QCD}\label{secII}

The AdS/CFT conjecture~\cite{Maldacena:1997re} offers one of the most promising ways to study gauge theories in strongly coupled regimes through their weakly coupled supergravity duals compactified in AdS$_5$. In the recent years attempts have been directed towards using the gauge/string duality to QCD, {\it{i.e.}}, a holographic equivalence is conjectured between four-dimensional strongly coupled QCD at large $N_c$ and a five-dimensional weakly interacting gauge theory coupled to gravity on a five-dimensional space not necessarily (but asymptotically) AdS$_5$. The fact that QCD is conformally invariant in the deep Euclidean makes the gauge/string duality a good starting point towards a theory of hadrons. However, QCD is not conformal in the strongly coupled regime and crucial ingredients like confinement or chiral symmetry breaking have to be incorporated. Confinement can be easily modeled by making the bulk space compact, for instance by placing an infrared brane some distance $z_0$ away from the ultraviolet brane. The QCD resonances are then the Kaluza-Klein modes arising from the compactification. Chiral symmetry breaking is more involved but AdS/CFT seems to have the potential to describe both explicit and spontaneous symmetry breaking~\cite{Klebanov:1999tb}. The pion field can also be incorporated, and the agreement of the whole picture with QCD, especially with vector mesons, is quite remarkable~\cite{Erlich:2005qh,DaRold:2005zs,Sakai:2004cn,Sakai:2005yt}.

In order to bridge the gap between the original AdS/CFT conjecture and AdS/QCD two main approaches have been followed, the so-called top-down and bottom-up. In the top-down approach, one looks for a suitable setting of D-branes in string theory, with gauge theory on their world volume, which at low energy would
produce an effective background geometry for the dual five-dimensional gravitational theory. An example is the Sakai-Sugimoto model~\cite{Sakai:2004cn} to be considered later.

The bottom-up approach is more phenomenologically oriented and one starts directly from warped five-dimensional models, with an AdS$_5$ metric in the ultraviolet regime but with drastic deviations
in the infrared, where nonperturbative effects of QCD force a description in terms of new low-energy degrees of freedom. In this paper we will consider Hard-Wall (HW) and Soft-Wall (SW) models. Both in
the HW models of~\cite{Erlich:2005qh} and~\cite{Hirn:2005nr}, the AdS$_5$ space is cut off at a finite size, producing an infinite number of Kaluza-Klein resonances, to be identified with the hadronic spectrum. The two models differ in the implementation of $\chi$SB, though. In~\cite{Erlich:2005qh}, $\chi$SB is induced by a 5D scalar field, holographically dual of the $\bar q q$ operator of QCD, whose nonvanishing vacuum expectation value is responsible for
$\chi$SB. In contrast, in~\cite{Hirn:2005nr}, $\chi$SB is achieved by imposing
appropriate infrared boundary conditions. The SW model
proposed in~\cite{Karch:2006pv} is a five-dimensional holographic model in which the AdS$_5$ space is noncompact. Confinement and an infinite number of bound states follow from the presence of a nontrivial dilaton background. The main feature of the SW model is its ability to produce an hadronic spectrum with Regge behavior, which leads to better agreement with the resonance spectrum in QCD.

In any of the models above, one boldly conjectures the applicability of the holographic recipe to compute correlation functions of the dual 4-dimensional theory~\cite{Gubser:1998bc,Witten:1998qj}. For every quantum operator ${\cal O}(x)$ in QCD, there exists a corresponding bulk field $\phi(x,z)$, whose value on the ultraviolet brane, $\phi(x,0)\equiv\phi_0(x)$, is identified with the four-dimensional source of ${\cal O}(x)$. Hence, the generating functional of the four-dimensional theory can be computed from the five-dimensional action evaluated \emph{on-shell} (neglecting stringy corrections), {\it{i.e.}},
\begin{equation}
\mbox{exp}\left(i S_5[\phi_0(x)]\right)=\langle\mbox{exp}\left[i\int d^4
x \,\phi_0(x)\,{\cal O}(x)\right]\rangle_{\rm QCD_4}~.\label{holography}
\end{equation}
Integrating by parts the quadratic part of the five-dimensional action on-shell effectively reduces to a boundary four-dimensional term quadratic in $\phi_0(x)$. By varying the action with respect to $\phi_0(x)$ one can generate the different connected $n$-point Green's functions of QCD.

The simplest five-dimensional action can be generically written in the form
\begin{equation}\label{S5}
S_5=S_{\rm YM}+S_X+S_{\rm CS}~,
\end{equation}
where
\begin{widetext}
\begin{eqnarray}\label{action}
S_{\rm YM}&=&-{\rm tr}  \int d^4x \int_0^{z_0} dz~e^{-\Phi(z)} \frac{1}{8g_5^2}w(z)\left[{\cF}_{(L)}^{MN}{\cF}_{(L)MN}+{\cF}_{(R)}^{MN}{\cF}_{(R)MN}\right]~,\nonumber\\
S_X&=&{\rm tr}  \int d^4x \int_0^{z_0}
 dz~e^{-\Phi(z)}w(z)^3\left[D^{M}X D_{M}X^{\dagger} +
V (X^{\dagger}X)\right]~,\nonumber\\
\end{eqnarray}
\end{widetext}
with $\cF_{MN} = \partial_{M}\cB_{N} -
\partial_{N} \cB_{M}-i[{\cal B}_{M}, {\cal B}_{N}]$ and $\cB_{L,R} = V \mp A $, where $V(A) \in U(2)_{V(A)}$ are vector (axial-vector) five-dimensional fields. If these are coupled to Dirac currents, then the holographic prescription predicts that they are massless and hence gauge-invariant. It is common to work in the axial gauge $V_5=A_5=0$. The surviving five-dimensional gauge fields $V_{\mu}(x,z)$ and $A_{\mu}(x,z)$ holographically correspond to four-dimensional vector and axial-vector QCD currents, $\bar q \gamma_\mu q$ and $\bar q_R \gamma_\mu \gamma_5 q_R$, respectively.

$\Phi(z)$ is a dilaton field and $X(x,z)$ a scalar field transforming under the chiral group as $g_L X g_R^{\dagger}$. Then, accordingly, $D_M X =
\partial_M X - iL_M X + iX R_M$. As we shall see below, the presence of the scalar field $X$ and the form of its potential depend on the model considered. For instance, in models where $\chi$SB is induced by chirally-asymmetric boundary conditions on the gauge fields, $X(x,z)$ is not essential.

The extra-dimension is taken to extend over the  interval $(0,z_0)$, where the upper limit can be infinite for some models. The metric of the five-dimensional space can be written generically in terms of a warp factor $w(z)$ as
\begin{align}
g_{MN}dx^Mdx^N = w(z)^2\left(\eta_{\mu \nu}dx^{\mu}dx^{\nu} -
dz^2\right)~,\label{metricwarp}
\end{align}
where $\eta_{\mu\nu}={\rm Diag}\,(1,-1,-1,-1)$, $\mu,\nu=(0,1,2,3)$ and $M,N=(0,1,2,3,z)$. In AdS$_5$ space the warping factor takes the form $w(z)=1/z$.

As pointed out in~\cite{Hill:2006wu}, anomalous processes in four dimensions can be reproduced from the five-dimensional Chern-Simons term
\begin{equation}
S_{\rm CS}[\cB]=\frac{N_c}{24\pi^2} \int
\tr\left( \cB
\cF^2-\frac{i}{2}\cB^3\cF-\frac{1}{10}\cB^5 \right)~.
\end{equation}
In order to account for chirality, one should work with
\begin{align}
S^{\mathrm{AdS}}_{\mathrm{CS}}[\mathcal{B}_L, \mathcal{B}_R] = S_{%
\mathrm{CS}}[\mathcal{B}_L] - S_{\mathrm{CS}}[\mathcal{B}_R]~.\label{omega5}
\end{align}
In the following we will briefly review the general features of the holographic
models we will consider for our analysis of the electromagnetic pion form factor, mainly to fix our notation. Further details can be found in the original literature.


\subsection{Hard-Wall models}
The distinguishing features of HW models are a compact fifth dimension, $0\leq z\leq z_0$ and a constant dilaton field $\Phi(z)$. We will concentrate on two such models, which we name HW1
~\cite{Erlich:2005qh,DaRold:2005zs} and HW2~\cite{Hirn:2005nr}. The presence of the upper bound $z=z_0$ defines an infrared brane, producing an explicit breaking of the scale invariance of the dual four-dimensional theories at energies $\approx 1/z_0\approx 1$ GeV. Moreover, in the
presence of this cutoff,  Wilson loops follow an area-law behavior in the infrared regime, thereby simulating the onset of a confining phase.

In compliance with $\chi$SB in the four-dimensional theory, axial gauge invariance is broken in the five-dimensional background, the longitudinal component of the axial-vector field becoming physical and related to the pion field.

HW1 and HW2 differ in the mechanism leading to the $\chi$SB. In the HW1 model, the breaking is due to the scalar field $X(x,z)$, whose coupling to the axial-vector gauge field produces an effective mass term, breaking the gauge invariance in the axial sector. In the HW2 model, instead, no such
field is present in the five-dimensional Lagrangian (\ref{S5}) and $\chi$SB is achieved through different infrared boundary conditions for vector and axial-vector fields.


\subsubsection{$\chi$SB from a scalar bulk field: HW1}
One considers the action (\ref{action}) without dilaton field and with the complex scalar field $X(x,z)=
v(z)U(x,z)/2$, where $U$ contains the pion field and $v(z)$ is the scalar component that breaks chiral symmetry in the bulk. The scalar potential is reduced to the mass term, given by:
\begin{equation}
\label{massX} V(X^{\dagger}X)= \frac{3}{z^5} X^{\dagger}X~.
\end{equation}
The previous expression guarantees that $X$ can be coupled to ${\bar{q}}q$. Solving the five-dimensional equations of motion at zero four-dimensional momentum one finds $v(z) = (m_q z + \sigma z^3) $, where the parameters $m_q$ and $\sigma$ are holographically identified as the quark mass and the $\bar q q $ condensate, \emph{i.e} the sources of explicit and spontaneous $\chi$SB, respectively.

A nonvanishing $ v(z)$ induces a $z$-dependent mass term for the axial-vector field in Eq.~(\ref{action}), thereby breaking the degeneracy between vector and axial-vector resonances. The scalar field $X$ also produces a nontrivial coupling between the longitudinal component of the axial-vector field and the pion field. Defining $A_{\mu \,\parallel}^a(q,z)= -iq_\mu \varphi(q,z)$ and $U(x,z)=\exp{\left [2i t^a \pi^a(x,z)\right]}$ (for $SU(2)$, $t^a=\sigma^a/2$, with $\sigma^a$ being the Pauli matrices) one gets the system of coupled equations~\cite{Erlich:2005qh}:
\begin{eqnarray}
\partial_z\left(\frac{1}{z} \, \partial_z \varphi^a \right)
+\frac{g_5^2 \, v(z)^2 }{z^3} (\pi^a-\varphi^a)&=&0~,\label{HW-AL}\\
-q^2\partial_z\varphi^a+\frac{g_5^2 \, v(z)^2}{z^2} \, \partial_z \pi^a &=&0~.\label{HW-Az}
\end{eqnarray}

As a result, the HW1 model satisfies the Gell-Mann--Oakes--Renner (GMOR) relation~\cite{GellMann:1968rz}:
\begin{equation}
m_\pi^2f_\pi^2=(m_u+m_d)<\bar{q}q>=2m_q\sigma~.
\end{equation}
The pion wave function $\varphi(z)=1-\Psi(z)$ can be obtained from the coupled equations~(\ref{HW-AL})
and (\ref{HW-Az}) which describe the dynamics in the axial sector. In the chiral limit, $m_q=0$, one obtains
\begin{align}
\Psi(z)=\Gamma[2/3]&\left(\frac{\xi z^3}{2}\right)^{\frac{1}{3}}\left[
I_{-1/3}(\xi z^{3})-\right.\nonumber\\
&\left.-\frac{I_{2/3}(\xi z_0^{3})}{I_{-2/3}(\xi z_0^{3})}I_{1/3}(\xi z^{3})\right]\label{pionHW1}~,
\end{align}
where $\xi \equiv g_{5}\sigma /3$, $\sigma $ being the quark condensate. The pion wave function
is different from zero on the infrared brane,
\begin{equation}\label{pionIR}
\Psi (z_0)=\frac{\sqrt{3}\Gamma[2/3]}{\pi I_{-2/3}(\xi z_0^3)}
\left(\frac{1}{2(\xi z_0^3)^2}\right)^{1/3}~,
\end{equation}
a fact that will be important in the evaluation of the anomalous amplitude.

By solving the equation of motion for the vector field ($Q^2=-p^2$ being the 4-dimensional Euclidean momentum):
\begin{equation}\label{HW-V}
\partial_z\left(\frac{1}{z} \, \partial_z \cJ \right)-\frac{Q^2 }{z} \cJ = 0~,
\end{equation}
subject to the (Dirichlet) ultraviolet boundary condition $\cJ(Q,0)=1$ and the (Neumann) infrared one, $\partial_y\cJ(Q,z_0)=0$, one finds the so-called vector bulk-to-boundary propagator $\cJ(Q,z)$, which can be written in terms of Bessel functions:
\begin{equation}
{\cal J}(Q,z)={Qz}\left[K_1(Qz)+I_1(Qz)\frac{K_0(Qz_0)}{I_0(Qz_0)}\right]~.
\end{equation}\label{JQz}
Vector resonances are associated with solutions $\psi_n(z)$ of
the equation of motion for the vector field at discrete $p^2=m_n^2$, with vanishing boundary conditions: $\psi_n(0)=0$, and $\partial_z\psi_n(z_0)=0 $.


\subsubsection{$\chi$SB through boundary conditions: HW2}
The action of the HW2 model \cite{Hirn:2005nr} is entirely given by the
gauge field part of eq.(\ref{action}). The role of the scalar field $X$ as the source of chiral symmetry breaking is played by asymmetric boundary conditions between vector and axial fields. In this model, vectors are required to obey infrared Neumann boundary conditions while axial fields satisfy Dirichlet boundary conditions. This leads to a splitting in the mass spectra that qualitatively reproduces what is observed in nature.

The pion field is built from Wilson lines extending between the 5D boundaries:
\begin{align}
U(x)=\xi_{R}(x)\xi_{L}^{-1}(x)~,\label{HW2U}
\end{align}
where
\begin{equation}
\xi_{L,R}(x)= P  \exp \left\{-i \int_0^{z_0} dz'\, \cB^{L,R}_z(x,z')
\right\}~.
\end{equation}
The fact that the pion is a nonlocal object leads to difficulties to implement the GMOR relation. For our purposes, this will not be of relevance, since we shall consider the HW2 model only in the chiral
limit of massless pions.

Chiral symmetry breaking is implemented by splitting the fields like
\begin{align}
\hat{V}_{\mu}\left(x, z\right) &\equiv V_{\mu}\left(x,z\right) +
\hat{V}_{\mu} \left(x,0\right)~, \\ \nonumber
\hat{A}_{\mu}\left(x,z\right) &\equiv  A_{\mu}\left(x,z\right) +
\alpha\left(z\right)\hat{A}_{\mu}\left(x,0\right)~,
\end{align}
where the last terms are the sources and $\alpha \left(z\right)$, which plays the role of the pion wave function, is determined by demanding no mixing between the pion and the axial resonances:
\begin{eqnarray}
\alpha\left(z\right)&=&1-\frac{z^2}{z_0^2}~.
\end{eqnarray}
For comparison, we will also consider the flat case, {\it{i.e.}} $w(z)=1$. In that case the bulk-to-boundary propagator takes the form
\begin{equation}
\cJ(Q,z)=\cosh(Qz)+\tanh(Qz_0)\sinh(Qz)~,
\end{equation}
and the pion wave function is given by
\begin{equation}
\alpha(z)=1-\frac{z}{z_0}~.
\end{equation}


\subsubsection{The Sakai-Sugimoto as an HW2 model}
The original action of \cite{Sakai:2004cn} is
\begin{equation}
S=S_{\rm YM}+S_{\rm CS}~,\label{SSmodel}
\end{equation}
with
\begin{equation}
S_{\rm YM}=-\kappa \int d^4 x
\int_{-\infty}^{\infty}dz\,\tr\left[\,
\half\,h(z){\cF}_{\mu\nu}^2-k(z){\cF}_{\mu z}^2 \right]~,\label{SSmodel1}
\end{equation}
where the functions $h(z)$ and $k(z)$ are
given by
\begin{equation}
h(z)=(1+z^2)^{-1/3}~;\quad k(z)=1+z^2~, \label{hk}
\end{equation}
and $S_{\rm CS}$ is given in Eq.~(\ref{omega5}).

The constant $\kappa$ is related to the 't~Hooft
coupling $\lambda$ and the number of colors $N_c$ as
\begin{equation}
\kappa=\frac{\lambda N_c}{216\pi^3}~. \label{kappa}
\end{equation}
The model also has a mass scale $M_{KK}$ which in (\ref{SSmodel1}) was absorbed in the dimensionless parameter $z$. For our numerical analysis, the two parameters will be chosen as~\cite{Sakai:2004cn}
\begin{align}
M_{KK}=949\mbox{ MeV}~,~~~
\kappa=0.00745~,
\label{Mkkkappa}
\end{align}
in order to fit the experimental values of the $\rho$ meson mass, $m_\rho\simeq 776$ MeV, and the pion decay
constant, $f_\pi\simeq 92.4~{\rm MeV}$.

The action (\ref{SSmodel}) was obtained in \cite{Sakai:2004cn} as the effective action of $N_f$ probe D8-branes placed in the background of $N_c$ D4-branes, studied in~\cite{Witten:1998zw}, and stands for an effective theory of mesons in four-dimensional (large $N_c$) QCD with $N_f$ massless quarks. In the following we will reformulate the Sakai-Sugimoto model and show that it can be cast in the form of a HW2 model. This reformulation will prove useful for computational purposes in the following Section.

As a first step, let us define $y=\tan^{-1} z$, with values in the
finite interval $-\pi/2\leq y \leq \pi/2$, whose end points $y=\pm
\pi /2 $ correspond to the ultraviolet branes while the point $y=0$
corresponds to the infrared brane. In the $A_z=0$ gauge, (\ref{SSmodel1}) becomes
\begin{equation}
S_{\rm YM}=-\kappa \int d^4 x
\int_{-\pi/2}^{\pi/2}dy\,\tr\left[\,\half \widetilde{h}(y)
{\cF}_{\mu\nu}^2-{\cF}_{\mu y}^2 \right]~,\label{modely}
\end{equation}
where $\widetilde{h}(y)$ is an {\emph{even}} function of $y$. This
allows us to decompose the gauge field $\cA_\mu$ in parity-even and
parity-odd parts as $\cA_\mu=V_\mu+A_\mu$, with $V_\mu (x,-y)=V_\mu(x,y)$ and $A_\mu(x,-y)=-A_\mu(x,y)$, and restrict the theory to half the interval, {\it{i.e.}} $0\le y\le \pi/2$. Parity transformations in the fifth dimension correspond to the exchange of left and right-handed chiralities and can be
used to distinguish between vectors and axial-vector fields (vectors and axial-vectors having even and odd $y$-profiles, respectively). As a consequence, they satisfy different boundary conditions on the infrared brane (at $y=0$), \emph{i.e.} Neumann for vectors, $\partial_y V_{\mu}|_{y=0}=0$, and Dirichlet $A_{\mu}|_{y=0}=0$ for axial-vectors.

Next, let us define the dimensionless variable $z\equiv \pi/2-y$, with $0\le z\le \pi /2$ (not to be confused with the original variable in (\ref{SSmodel1})). $z=0$ corresponds to the ultraviolet brane and $z=\pm \pi /2 $ to the infrared brane, and the action takes the following form:
\begin{eqnarray}
S_{\rm YM}&=& -\kappa  \int d^4 x \int_0^{\pi/2}dz \tr \Big\{ (\sin
z)^{-4/3}\nonumber\\
&&\!\!\!\!\!\!\!\!\!\!\times \left[(V_{\mu\nu}-i[A_{\mu},A_{\nu}])^2+(D_{\mu} A_{\nu}
-D_{\nu} A_{\mu})^2 \right] \nonumber\\
&& - 2(\partial_z V_{\mu})^2-2(\partial_z A_{\mu})^2 \Big\}~,
\label{actionz}
\end{eqnarray}
where $V_{\mu\nu}=\partial_{\mu} V_{\nu}-\partial_{\nu} V_{\mu}-i[V_{\mu} , V_{\nu}]$ and $D_{\mu} A_{\nu}=\partial_{\mu} A_{\nu}-i[V_{\mu} , A_{\nu}]$.

In the new variable $z\in[0, \pi /2]$ the SS model
takes the form of the HW2 model with different boundary conditions
for vector and axial-vector fields on the infrared boundary $z=\pi /2$,
and an effective metric which is  not AdS$_5$:
\begin{align}\label{metric}
g_{MN}dx^Mdx^N = (\sin z)^{-\frac{4}{3}}\eta_{\mu \nu}dx^{\mu}dx^{\nu} -
(\sin z)^{-\frac{8}{3}} dz^2~.
\end{align}
The chiral field containing the pion is also built from Wilson lines extending between boundaries. One can easily show that in the SS model the function $\alpha(z)$ discussed in the previous subsection takes the simple form
\begin{equation}
\alpha(z)=1-\frac{2z}{\pi}~.
\end{equation}

Vector (and axial-vector) fields can be obtained by solving the bulk-to-boundary differential equation
\begin{equation}
\left(\frac{d^2}{dz^2}-\frac{Q^2}{(\sin z)^{4/3}}\right)\cJ=0~,
\end{equation}
subject to the boundary conditions ${\cal J}_{V,A}(Q,0) = 1 $, and $
\partial_z{\cal J}_V(Q,\pi/2) = 0$, ${\cal J}_A(Q,\pi/2) = 0$, for which there is no analytic expression. In Fig.~1 we show the first three normalized eigenfunctions for
vector and axial-vector fields in the new extra-dimensional
variable $z$.
\begin{figure}[htbp]
\centering \epsfysize=5cm\epsfxsize=8cm\epsfbox{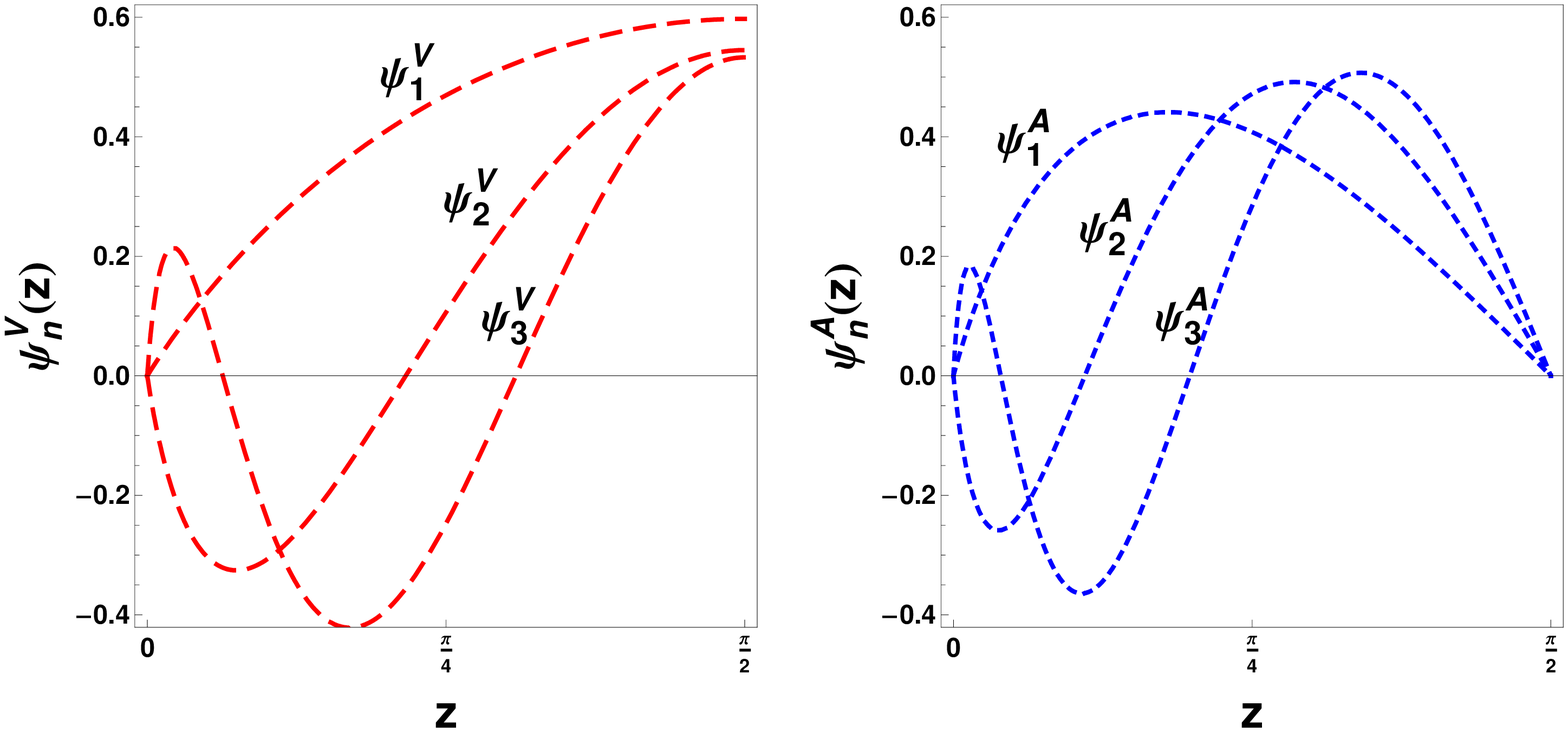}
{\caption{\small The first three vector and axial-vector meson
resonances of the SS model, using the new variable $z\in[0, \pi
/2]$. }}\label{allVect}
\end{figure}
As usual, the Fourier transform of the five-dimensional gauge field has been written as $\cA_\mu(q, z)=\widetilde \cA_\mu(q)\psi(z)$, with $\psi(z)$  satisfying the equation
\begin{equation}
\psi(z)''+\frac{q^2}{(\sin z)^{4/3}}\psi(z)=0 \label{eomz}~.
\end{equation}
Dynamical resonances $\psi_n(z)$ then correspond to normalized solutions of
(\ref{eomz}) with $q^2=m_n^2$, where the normalization condition is given by
\begin{equation}
4\kappa \int_{0}^{\pi/2}(\sin
z)^{-4/3}\psi_m(z)\psi_n(z)=\delta_{mn}~,\label{normalization}
\end{equation}
to ensure the canonical four-dimensional kinetic term for $\cA_\mu(x)$; the discrete spectrum of eigenvalues $q_n\equiv m_n$, with the corresponding eigenfunctions, can then be obtained numerically.


\subsection{Soft-Wall models}

SW models were originally motivated as holographic models with vector resonances displaying Regge trajectories. The action of SW models is given by (\ref{S5}), where the metric is AdS$_5$, and $e^{-\Phi(z)}$ represents a nontrivial
background dilaton field given by~\cite{Karch:2006pv}
\begin{equation}
\Phi(z)= \kappa^2 z^2~,
\end{equation}
leading to a spectrum given by
\begin{equation}
m_n^2=4\kappa^2(n+1)~.
\end{equation}
The dimensionful constant $\kappa$ can be fixed by fitting the mass of the first vector resonance to that of the $\rho$ meson, \emph{i.e.} $\kappa=m_\rho/2$. Therefore, the equations of motion for the different fields can be obtained from those of the HW1 model by replacing the AdS$_5$ warp factor with $e^{-\Phi(z)}/z$. Contrary to the HW models, in the SW model the extra dimension is no longer restricted to a finite interval, \emph{i.e. $z_0=\infty$} and the infrared boundary conditions are replaced by
normalization conditions on wave functions and bulk-to-boundary propagators.

For the vector fields one has to solve the following equation:
\begin{equation}\label{SW-V}
\partial_z\left(\frac{e^{-\kappa^2z^2}}{z} \, \partial_z \cJ \right)
 - Q^2\frac{e^{-\kappa^2z^2}}{z} \cJ = 0~,
\end{equation}
whose solution can be cast in the integral representation~\cite{Grigoryan:2007my}
\begin{align}
{\cal{J}}(Q,z)=\kappa^2z^2\int_0^{1}\frac{x^a}{(1-x)^2}\,\exp\left[-\frac{x}{1-x}\,\kappa^2z^2\right]\,dx~,\label{SWvectorbtb}
 \end{align}
where $a= Q^2/4
\kappa^2$.

One of the main drawbacks of the orginal SW is that chiral symmetry breaking is not implemented in a satisfactory way. Due to the absence of an infrared brane, the parameters for explicit and spontaneous symmetry breaking are not independent. This not only invalidates general relations like the GMOR, but makes the whole pion dynamics unclear. Recent proposals have tried to circumvent this problem, so far only at an heuristic level. In this paper we will adopt the prescription given in~\cite{Grigoryan:2008up}, where the pion wave function is assumed to be Gaussian,
\begin{align}\label{gaussian}
\alpha(z) =e^{-\kappa^2z^2}~.
\end{align}
A comparison of the pion wave function profiles for the different models is shown in Fig.~2.
\begin{figure}[h]
\centering \epsfysize=4cm\epsfxsize=6cm%
\epsfbox{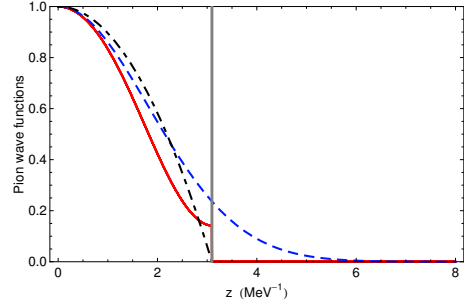} {\caption{\small Comparison of the pion
wave functions of the HW1 model (red solid line), the HW2 model (black dot-dashed line) and the Gaussian ansatz of Eq.~(\ref{gaussian}) (blue dashed line). The vertical line corresponds to the infrared brane at $z=z_0$.}}\label{PloFinale}
\end{figure}


\section{$\pi^0 \gamma ^* \gamma^* $ form factor}\label{secIII}

In this Section we will apply the methods described previously for the anomalous pion form factor. We will define the $\pi^0 \gamma^* \gamma^* $ form factor as
\begin{align}
\int d^4 x \ e^{iq_1\cdot x }& \langle 0|T\left\{J^{\mu }_{\mathrm{EM}%
}(x)\,J^{\nu}_{\mathrm{EM}}(0)\right\}| \pi^0(p) \rangle \\
&= \epsilon^{\mu \nu \alpha \beta}q_{1 \, \alpha} q_{2\, \beta} \,
F_{\gamma^*\gamma^*\pi^0} \left(Q_1^2,Q_2^2 \right ) \ ,  \nonumber
\end{align}
where $p = q_1 + q_2 $ is the pion momentum, $q_{1}, q_{2} $ are the momenta
of photons, and $q^2_{1,2} = -Q^2_{1,2} $. Notice that our conventions differ by a sign from the ones used in~\cite{Grigoryan:2007wn,Grigoryan:2008cc}.

Even though the form factor cannot be computed from first principles, there are certain kinematical limits where theoretical or experimental information is available. For instance, when both photons are on-shell, the form factor is determined (in the chiral limit) solely by the WZW anomaly term
\begin{equation}
F_{\gamma^*\gamma^*\pi^0} \left(0,0 \right )=-\frac{N_C}{12\pi^2f_{\pi}}~,
\end{equation}
and it is therefore convenient to define
\begin{equation}
F_{\gamma^*\gamma^*\pi^0} \left(Q_1^2,Q_2^2 \right
)=-\frac{N_C}{12\pi^2f_{\pi}}K(Q_1^2,Q_2^2)~,\label{FandK}
\end{equation}
such that $K(0,0)=1$.

On the experimental side, studies have focused on the low-energy behavior, when one of the photons is exactly on-shell and the other slightly off-shell. In this kinematic regime it is common to define the slope of the anomalous form factor $a_{\pi}$ as
\begin{eqnarray}
K(0,Q^2)&=&K(Q^2,0)\nonumber\\
&=&\left[1-\frac{a_{\pi}}{m_{\pi}^2}Q^2+\frac{b_{\pi}}{m_{\pi}^4}Q^4\cdots\right]~,
\end{eqnarray}
{\cal{i.e.}},
\begin{align}
a_{\pi} &= -m^2_{\pi}\left[\frac{d K(Q^2,0)}{dQ^2}\right]_{Q^2=0}~.
\end{align}
The world average is presently $a_{\pi}=0.032(4)$, mainly driven by the results of the CELLO Collaboration~\cite{Behrend:1990sr} on $\pi^0\to e^+e^-\gamma$. Given the importance of this kinematical regime for the evaluation of the $(g-2)_{\mu}$, we also define the curvature $b_{\pi}$ as
\begin{align}
b_{\pi} &= m^4_{\pi}\left[\frac{d K(Q^2,0)}{dQ^4}\right]_{Q^2=0}~.
\end{align}
Later on we will use the different holographic models to obtain a determination for both $a_{\pi}$ and $b_{\pi}$.

Another source of information comes when one of the photons is on-shell and the other far off-shell. Then the expected behavior of the pion form factor is dictated by the Brodsky-Lepage quark-counting rules~\cite{Lepage:1980fj,Brodsky:1981rp},
\begin{equation}\label{BL}
\lim_{Q^2\to\infty}K(0,Q^2)\sim \frac{1}{Q^2}~.
\end{equation}

Data on this kinematical regime is available from $e^+e^-\to e^+e^-\pi^0$. The behavior of the previous equation is reasonably reproduced by CLEO~\cite{Gronberg:1997fj} but incompatible with recent data by BABAR~\cite{:2009mc}. Hopefully upcoming BABAR data with better statistics will help clarify the situation. While the discrepancy is surprising, we want to emphasize that the kinematical regime we are considering does not accept an OPE expansion and, therefore, that Eq.~(\ref{BL}) is not on the same footing as a short-distance constraint.

Short-distance constraints can be obtained when both photons have large and equal virtualities. In this case,
\begin{equation}\label{OPEQ}
\lim_{Q^2\to \infty}K(Q^2,Q^2)=\frac{8\pi^2f_{\pi}^2}{N_c}\frac{1}{Q^2}~.
\end{equation}

It is worth stressing that all the constraints and experimental information presented above refer to the pion being strictly on-shell. In our case, since we are working in the chiral limit, to a very good approximation $p^2=m_{\pi}^2\sim 0$. We will come back to the on-shellness of the pion when we discuss the pion exchange diagram in $(g-2)_{\mu}$. For the time being, we will concentrate on the pion form factor as predicted from the different holographic models.

\subsection{Holographic predictions}

In terms of holographic QCD, $K(Q_1^2,Q_2^2)$ can be obtained from the VVA terms of the Chern-Simons action. Working in the axial gauge, $\cB_z=0$, and gathering the relevant pieces trilinear in the fields one finds:
\begin{equation}
S^{\rm{AdS}}_{\rm{CS}} = \frac{N_c}{24\pi^2}\epsilon^{\mu\nu\rho\sigma} \mathrm{tr} \int d^4 x\, dz \left(\partial_z \cB_{\mu}\right)\biggl[\cF_{\nu \rho}\cB_{\sigma} + \cB_{\nu}\cF_{\rho \sigma} \biggr]~.
\end{equation}
Replacing $A_{\mu}=\partial_{\mu}\pi$ and taking the pion to be on-shell one ends up with
\begin{align}
S_{\mathrm{CS}}^{\mathrm{AdS}}& =\frac{N_{c}}{12\pi ^{2}}\epsilon ^{\mu \nu
\rho \sigma }\int d^{4}x\int_{0}^{z_{0}}dz\ \pi ^{a} \\
& \times \left[ 2\,\, \partial_{z}\beta\,\, \partial
_{\rho }V_{\mu }^{a}\,\, \partial _{\sigma }\hat{V}_{\nu
} -\beta
\partial _{z}\left( \partial _{\rho }V_{\mu }^{a}\ \partial _{\sigma }\hat{V}%
_{\nu }\right) \right]~,  \nonumber
\end{align}
where $\beta(z)$ stands for the pion wave function, which we will denote as $\Psi(z)$ or
$\alpha(z)$, depending on the model.

Integrating by parts over $z$ in the second term above and dismissing a boundary term (cf.~\cite{Grigoryan:2007wn}) in order to reproduce the standard four-dimensional WZW action one gets
\begin{equation}
S_{\mathrm{CS}}^{\mathrm{AdS}}=\frac{N_{c}}{4\pi ^{2}}\epsilon^{\mu\nu\rho\sigma}\!\!\int_{0}^{z_{0}}\!\!dz\,(\partial _{z}\beta)\!\!\int d^{4}x\ \pi
^{a}\left( \partial _{\rho }V_{\mu }^{a}\right) \left( \partial _{\sigma }%
\hat{V}_{\nu }\right)~.
\end{equation}

Variation of  $S_{\mathrm{CS}}^{\mathrm{AdS}}$ above gives the three-point function:
\begin{equation}
T_{\alpha \mu \nu }(p,q_{1},q_{2})=\frac{N_{c}}{12\pi ^{2}}\frac{p_{\alpha }%
}{p^{2}}\,\epsilon _{\mu \nu \rho \sigma }\,q_{1}^{\rho }q_{2}^{\sigma
}K(Q_{1}^{2},Q_{2}^{2})~.
\end{equation}
In compliance with the large-$N_c$ limit, in extra-dimensional models the form of the interaction vertex between the pion and two external electromagnetic currents is mediated by the full tower of Kaluza-Klein vector mesons. From the variation of the Chern-Simons action with respect to the external sources one
obtains
\begin{eqnarray}\label{rec}
K(Q_1^2,Q_2^2)&=&-\int_{0}^{z_{0}}\cJ(Q_1,z)\cJ(Q_2,z)\,\partial _{z}\beta (z)\,dz\,\nonumber\\
&&+\,\,\mathrm{(possible\,\,boundary\,\,terms)}~,
\end{eqnarray}
where $\mathcal{J}(Q,z)$ is the vector bulk-to-boundary propagator
at Euclidean momentum $Q^{2}=-q^{2}$, defined in the previous
Section.

The significance of boundary terms in the Chern-Simons action was studied in detail in~\cite{Hill:2006wu}. In the context of the HW1 model, boundary terms in the expression of $K(Q_{1}^{2},Q_{2}^{2})$ were shown to be needed in order to have the right normalization required by the QCD axial anomaly, namely $K(0,0)=1$. The presence of these boundary terms is closely related to the infrared behavior of the pion wave function. As pointed out in Eq.~(\ref{pionIR}), in the HW1 model the value of the pion wave function at $z_{0}$ does not cancel and requires the addition of a boundary
term~\cite{Grigoryan:2007wn}
\begin{eqnarray}\label{rec1}
K(Q_1^2,Q_2^2)&=&-\int_{0}^{z_{0}}\cJ(Q_1,z)\cJ
(Q_2,z)\,\partial _{z}\Psi (z)\,dz\, \nonumber \\
&& +\cJ(Q_{1},z_{0})\cJ(Q_{2},z_{0})\,\Psi (z_0)~.
\end{eqnarray}

No boundary term is needed both in the case of the HW2 and the
SS models. The correct normalization is obtained in both cases (using $
\cJ(0,z)=1$) due to the boundary conditions satisfied by $\alpha(z)$:
\begin{equation}
K(0,0)=-\int_{0}^{z_{0}}\partial _{z}\alpha (z)\,dz=\alpha (0)=1~.
\end{equation}

For the SW models, since the pion wave function is introduced by hand, it depends on the ansatz made. In general, pion wave function that do not cancel at the infrared brane will require boundary terms to correctly implement the axial anomaly in $K(Q_1,Q_2)$. 

Equipped with Eqs.~(\ref{rec}) and (\ref{rec1}), together with the different expressions for the pion wave function and vector bulk-to-boundary propagator, one can easily find the expressions for the pion form factor. We are mostly interested in checking whether short-distance constraints are satisfied and to what extend the low-energy information complies with experimental data.

\subsection{Large-$Q^2$ behavior}

In asymptotically AdS holographic  models, it is easy to show that the function $K(Q_1,Q_2)$ automatically satisfies the high-energy constraints discussed in the previous Section. For instance, for arbitrarily large $Q_1$ and $Q_2$, and working in the HW2 model, it is not difficult to show that the general expression reads
\begin{eqnarray}\label{asymp}
K(w,Q^2)&\simeq&\frac{2}{z_0^2Q^2}\sqrt{1-w^2}\nonumber\\
&&\!\!\!\!\!\!\!\!\!\!\!\!\!\!\!\!\times\int_0^{\infty}d\xi\, K_1\left(\sqrt{1+w}\xi\right)K_1\left(\sqrt{1-w}\xi\right)\,\xi^3\nonumber\\
&\simeq&\frac{2}{w^3z_0^2Q^2}\left[w-(1-w^2)\tanh^{-1}{w}\right]~,\nonumber\\
\end{eqnarray}
where $w=\displaystyle \frac{Q_1^2-Q_2^2}{2Q^2}$, $2Q^2=Q_1^2+Q_2^2$ and $\xi=Qz$. It is easy to show that the momentum dependence of the previous result also holds for the HW1 and SW models: AdS dictates the large-$Q^2$ behavior of the bulk-to-boundary propagator, and this is common to all of them. Moreover, the shape of the pion wave function is very similar for all the models close to the ultraviolet boundary, as illustrated in Fig.~2. Indeed, as pointed out in \cite{Grigoryan:2008cc}, $\alpha(z)$ and the pion wave function (\ref{pionHW1}) coincide in the deep ultraviolet, since at small $z$
\begin{equation}
\partial_z\Psi(z)\simeq-f_{\pi}^2g_5^2z=-2\frac{z}{z_0^2}=\partial_z\alpha(z)~.
\end{equation}
Notice that differences between HW1 and HW2 only affect the infrared of the theory, and therefore are of no relevance for the large-$Q^2$ behavior (at most exponentially suppressed like $\mathcal{O}(e^{-Q z_0})$).

The form of the coefficient in front of Eq.~(\ref{asymp}) will obviously change depending on the details of the model. For instance, the SW model with the {\it{ad hoc}} Gaussian pion wave function is recovered by replacing $z_0\to 1/\kappa$. Since $z_0$ is related to the pion decay constant, while $\kappa$ is matched to the $\rho(770)$ mass, if we impose numerical agreement between both models we find the relation
\begin{equation}
m_{\rho}^2=8\pi^2f_{\pi}^2~,
\end{equation}
which can be compared for instance with the prediction~\cite{Golterman:1999au}
\begin{equation}
m_{\rho}^2=\frac{16\sqrt{6}}{5}\pi^2f_{\pi}^2~,
\end{equation}
coming from a large-$N_c$ sum rule analysis of $\Pi_{VV}$ and $\Pi_{AA}$. It is reassuring that both predictions are in excellent agreement.

When $Q_1=Q_2=Q$, Eq.~(\ref{asymp}) simplifies to
\begin{equation}\label{equal}
K(Q^2,Q^2)=\frac{4}{3z_0^2Q^2}~.
\end{equation}
With the expression for $z_0$ in the HW2 model, $z_0=2/g_5^2 f_\pi$, and $g_5^2=12\pi^2/N_c$, it is easy to show that it agrees with Eq.~(\ref{OPEQ}). Actually, one can go even further to show that the leading terms for $\lambda \rightarrow\infty$ are
\begin{align}\label{KNOPE}
K(\lambda^2 Q^2, & (\lambda^2 Q^2-P^2))\simeq \nonumber
\\
& \frac{2}{3}\frac{g_5^2 f_\pi^2}{Q^2}\left\{\frac{1}{\lambda^2}+
\frac{1}{\lambda^3}\frac{P\cdot
Q}{Q^2}+\mathcal{O}\left(\frac{1}{\lambda^4}\right)\right\}~,
\end{align}
which, up to $\mathcal{O}(\alpha_S)$
corrections which cannot be captured by the HW2 model, is the
short-distance behavior found in \cite{Knecht:2001qf} and
\cite{Knecht:2001xc}.

One can also explore the regime where one photon is on-shell and the other far off-shell. In that case,
\begin{equation}\label{BLrel}
K(0,Q^2)\simeq \frac{2}{z_0^2Q^2}\int_0^{\infty}d\xi\, K_1(\xi)\xi^2=\frac{4}{z_0^2Q^2}=\frac{8\pi^2f_{\pi}^2}{Q^2}~,
\end{equation}
which displays the Brodsky-Lepage scaling, {\it{cf.}} Eq.~(\ref{BL}).

Eqs.~(\ref{asymp}),~(\ref{equal}),~(\ref{KNOPE}) and (\ref{BLrel}) do not hold, however, for the models without asymptotic AdS metric, {\it{i.e}} the Sakai-Sugimoto and the flat HW2 model. At least for the latter, calculations can be performed analytically with the results
\begin{eqnarray}
K(w,Q^2)&=&\frac{1}{2wz_0Q}\nonumber\\
&&\!\!\!\!\!\!\!\!\!\!\!\!\!\!\!\!\!\!\!\!\!\!\times\big[\sqrt{1+w}\tanh{(Qz_0\sqrt{1+w})}-\{w\to-w\}\big]~,\nonumber\\
K(Q^2,Q^2)&=&\frac{1}{2Qz_0}~,\nonumber\\
K(0,Q^2)&=&\frac{1}{Qz_0}~,
\end{eqnarray}
which fail to reproduce the OPE of QCD.

\subsection{Small-$Q^2$ behavior and predictions for the parameters $a_{\pi}$ and $b_{\pi}$}
\bigskip
At small virtualities one can expand $K(Q_1^2,Q_2^2)$ in the form
\begin{eqnarray}\label{low}
K(Q_{1}^{2},Q_{2}^{2})\simeq 1+{\hat{\alpha}}~(Q_1^2+Q_2^2)+{\hat{\beta}}~ Q_1^2Q_2^2&&\nonumber\\
+{\hat{\gamma}}~(Q_1^4+Q_2^4)~.&&
\end{eqnarray}
By comparison with the previous Section, we immediately conclude that $a_{\pi}=-{\hat{\alpha}}m_{\pi}^2$ and $b_{\pi}={\hat{\gamma}}m_{\pi}^4$. The parameter ${\hat{\beta}}$ does not contribute to processes when one of the photons is real. Even so, it will be a useful parameter in the determination of $(g-2)_{\mu}$ in Section~\ref{secIV}.

Eq.~(\ref{low}) can be reproduced from holographic models by working out the small-$Q^{2}$ behavior of $\mathcal{J}(Q,z)$, which we will parametrize as
\begin{equation}
\mathcal{J}(Q,z)\equiv 1-Q^{2}g(z)+Q^4h(z)~.
\end{equation}
The functions $g(z)$ and $h(z)$ can be easily obtained by solving perturbatively in $Q^2$ the equation of motion for the different models. This leads to the following analytic expressions:
\begin{itemize}
\item[(i)] HW1 and HW2 models:
\begin{eqnarray}
g(z)&=&\frac{z^2}{4}\left[1-2\log\left(\frac{z}{z_0}\right)\right]~,\nonumber\\
h(z)&=&\frac{z^4}{16}\left[2\left(\frac{z_0}{z}\right)^2-\frac{5}{4}+\log\left(\frac{z}{z_0}\right)\right]~;
\end{eqnarray}
\item[(ii)] SS model:
\begin{eqnarray}
g(z)&=&{M_{KK}^{-2}}\left[ \int_{0}^{z}dy\frac{y}{(\sin y)^{4/3}}
+z\!\!\int_{z}^{\pi /2}\!\!\!\!\!\!\frac{dy}{(\sin y)^{4/3}}\right]~,\nonumber\\
h(z)&=&{M_{KK}^{-2}}\left[ \int_{0}^{z}dy\frac{y\,\, g(y)}{(\sin y)^{4/3}}
+z\!\!\int_{z}^{\pi /2}\!\!\!\!\!\!dy\frac{g(y)}{(\sin y)^{4/3}}\right]~;\nonumber\\
\end{eqnarray}
\item[(iii)] SW model:
\begin{eqnarray}
g(z)&=&-\frac{z^2}{4}\int_0^1 \exp \left[-\frac{x}{1-x}\kappa^2
z^2\right]\frac{\ln x}{(1-x)^2}dx~,\nonumber\\
h(z)&=&\frac{z^2}{32\kappa^2}\int_0^1 \exp \left[-\frac{x}{1-x}\kappa^2
z^2\right]\frac{\ln^2 x}{(1-x)^2}dx~;\nonumber\\
\end{eqnarray}
\item[(iv)] flat case:
\begin{eqnarray}
g(z)&=&z^2\left[\frac{z_0}{z}-\frac{1}{2}\right]~,\nonumber\\
h(z)&=&\frac{z^4}{24}\left[1-4\frac{z_0}{z}+8\left(\frac{z_0}{z}\right)^3\right]~.
\end{eqnarray}
\end{itemize}
\begin{table}[t]
\begin{tabular}{cccc}
\hline
Model &\,\,\,\,\, ${\hat{\alpha}}$ (GeV$^{-2}$) &\,\,\,\,\, ${\hat{\beta}}$ (GeV$^{-4}$)&\,\,\,\,\, ${\hat{\gamma}}$ (GeV$^{-4}$)\\
\hline
HW1 &\,\,\,\,\, -1.60 &\,\,\,\,\, 3.01&\,\,\,\,\, 2.63 \\
HW2 (AdS) &\,\,\,\,\, -1.81 &\,\,\,\,\, 3.65&\,\,\,\,\, 3.06 \\
HW2 (Flat) &\,\,\,\,\, -1.37 &\,\,\,\,\, 2.25 &\,\,\,\,\, 2.25\\
SS &\,\,\,\,\, -2.04 &\,\,\,\,\, 4.56&\,\,\,\,\, 3.55 \\
SW &\,\,\,\,\, -1.66 &\,\,\,\,\, 3.56&\,\,\,\,\, 2.76 \\
\hline
\end{tabular}
{\caption{Values of  ${\hat{\alpha}}$, ${\hat{\beta}}$ and ${\hat{\gamma}}$ for the holographic models discussed in the main text.}}\label{tab1}
\end{table}
Plugging the previous expressions into Eq.~(\ref{rec}) or (\ref{rec1}) one can obtain the different determinations for the low-energy parameters. For the HW1 model one finds
\begin{eqnarray}  \label{alfabetaHW1}
{\hat{\alpha}}&=&-\frac{z_0^2}{4} \Psi(z_0)+ \int_{0}^{z_0}g(z)\partial _{z}\Psi(z)dz~,\nonumber\\
{\hat{\beta}}&=&\frac{3z_0^4}{64} \Psi(z_0)- \int_{0}^{z_0}h(z)\partial
_{z}\Psi(z)dz~,\nonumber\\
{\hat{\gamma}}&=&\frac{z_0^4}{16} \Psi(z_0)- \int_{0}^{z_0}g(z)^2\partial
_{z}\Psi(z)dz~,
\end{eqnarray}
while for the remaining models:
\begin{eqnarray}  \label{alfabetaHW2}
{\hat{\alpha}}&=&\int_{0}^{z_0}g(z)\partial _{z}\alpha(z)dz~,\nonumber\\
{\hat{\beta}}&=&-\int_{0}^{z_0}h(z)\partial _{z}\alpha(z)dz~,\nonumber\\
{\hat{\gamma}}&=&-\int_{0}^{z_0}g(z)^2\partial _{z}\alpha(z)dz~.
\end{eqnarray}
The values for the different models are collected in Table~I.
Experimentally, only the slope has been determined,
${\hat{\alpha}}=-1.76(22)$ GeV$^{-2}$~\cite{PDG}, which is
correctly reproduced by the HW1, HW2 and SW models. Notice that,
even though the slope is a genuine low-energy quantity, models
without an asymptotic AdS metric fail to reproduce the experimental
value. The quartic parameters ${\hat{\beta}}$ and
${\hat{\gamma}}$, using the phenomenologically acceptable HW1, HW2
and SW models, are predicted to be
\begin{eqnarray}
{\hat{\beta}}=3.33(32)\,{\mathrm{GeV}}^{-4}, \label{alphaholography}\\
{\hat{\gamma}}=2.84(21)\,{\mathrm{GeV}}^{-4}.\label{betaholography}
\end{eqnarray}
 Both holographic predictions will be used in the
evaluation of the HLBL contribution to the $(g-2)_{\mu}$.
\subsection{A comment on lowest meson dominance}
Before moving to the evaluation of the $(g-2)_{\mu}$, it is important to explore the resonance contributions to the results of Table~1. For this, we write the vector bulk-to-boundary propagator in its spectral decomposition
\begin{equation}
{\cal J}(z, Q)=\sum_{n=1}^\infty
\frac{f_n}{Q^2+m_n^2}\psi_n(z),\label{btbseries}
\end{equation}
where $m_n$ is the mass of the $n$th vector resonance and $f_n$ is related to its decay constant. Plugging this expression into Eq.~(\ref{rec}), one obtains the double series
\begin{equation}
K(Q_1^2, Q_2^2)\equiv\sum_{k,l=1}^\infty\frac{B_{kl}}{(Q_1^2+m_k^2)(Q_2^2+m_l^2)}~.\label{Kexpand}
\end{equation}
This expression is of the form expected in the large-$N_c$ limit, with the full tower of vector mesons propagating between the pion and the two photons.
\begin{table}[t]%
\begin{tabular}{cccccccccc}
\hline
 Model & \multicolumn{3}{c}{$\hat\alpha_n /\hat\alpha$}
& \multicolumn{3}{c}{$\hat\beta_n /\hat\beta$} &
\multicolumn{3}{c}{$\hat\gamma_n/ \hat\gamma$} \\
\hline
HW1 &1.20 & -0.18& -0.04& 1.10& -0.06&0.01 & 1.20& -0.22& 0.06\\
HW2 (AdS) &1.30 & -0.37& 0.06& 1.10& -0.11&0.01 & 1.30& -0.37& 0.08\\
HW2 (flat) &0.99 & 0.01& 0.00& 1.00& 0.00&0.00 & 1.00& 0.00& 0.00\\
SS &1.70 & -1.10& 0.49& 1.30& -0.34& 0.07 & 1.60& -1.10& 0.54\\
SW &0.75 & 0.14& 0.05& 0.87& 0.09&0.02 & 0.88& 0.09& 0.02\\
\hline
\end{tabular}
{\caption{Contribution to $\hat \alpha$, $\hat\beta$ and $\hat\gamma$ due to the first three vector-meson radial excitations. For ${\hat{\alpha}}_n$, each subcolumn \mbox{($n=1,2,3$)} contains the contribution of $k,l \leq n$ terms.
$\hat \beta_n$ and $\hat \gamma_n$ are defined analogously.}}
\end{table}
The contributions for $\hat\alpha$, $\hat\beta$ and $\hat\gamma$ from the first three vector radial excitations are reported in Table~II. The important point to notice is that lowest vector dominance is a feature common to all models (from the  practically $\rho$-dominance of the flat model to the more moderate behavior of the SS model). Incidentally, notice also that, with the exception of the SW model, the contribution of the first resonance tends to overshoot the total value and has to be compensated by negative contributions from higher-order resonances. Finally, let us remark that the representation of Eq.~(\ref{Kexpand}) is only reliable for moderate values of the resonance indices. Most likely Eq.~(\ref{Kexpand}) is, at most, an asymptotic expansion and, therefore, beyond a certain threshold it ceases to be meaningful.


\section{The hadronic light by light contribution to the $(g-2)_{\mu}$}\label{secIV}

With experimental accuracies at the $10^{-10}$ level, precise determinations of the HLBL contribution to the $(g-2)_{\mu}$ become of paramount importance. The problem is that genuine nonperturbative techniques are required and having theoretical uncertainties under control is certainly a challenging task. The HLBL contribution to the $(g-2)_{\mu}$ is depicted in Fig.~3. When internal momenta in the loop are high enough one can use perturbation theory, but a proper evaluation also requires low energies, {\it{i.e.}} Goldstone bosons and hadrons. The neutral pion exchange contribution, depicted in Fig~4, turns out to be the dominant piece ($a_{\mu}^{(\pi^0)}\sim 7\cdot 10^{-10}$) followed by the $\eta$ and $\eta^{\prime}$ contributions ($a_{\mu}^{(\eta,\eta^{\prime})}\sim 3\cdot 10^{-10}$). Quark and Goldstone loops, axial-vectors and scalars are also expected to contribute at the level of $(1-2)\cdot 10^{-10}$, but cancellations occur (scalars and Goldstone loops contribute negatively) and so the Goldstone boson exchange ends up collecting the bulk of the effect. In particular, this means that the naive large$-N_c$ counting works and single-resonance exchange is the dominant effect. This observation suggests that procedures based on the $1/N_c$ framework are suitable tools to address the problem.
\begin{figure}[t]
\begin{center}
\includegraphics[width=5cm]{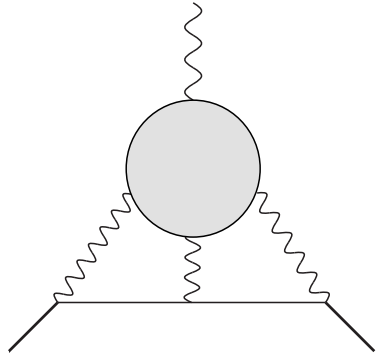}
\end{center}
\caption{Hadronic light by light contribution to $(g-2)_{\mu}$.\label{fig:2} }
\end{figure}

\begin{figure}[t]
\begin{center}
\includegraphics[width=5cm]{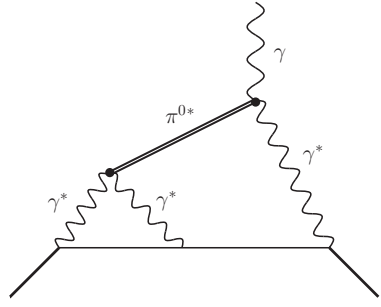}
\end{center}
\caption{One of the diagrams contributing to pion exchange in the HLBL contribution to $(g-2)_{\mu}$.\label{fig:21} }
\end{figure}

Different parametrizations have been used in the past to evaluate the HLBL contribution. In particular, there is the detailed study of~\cite{Knecht:2001qf} with different parametrizations based on LMD. Having in mind what we did so far, one could be tempted to use the results of the previous Section on the pion form factor to evaluate each of the $\pi^0\gamma\gamma$ vertices in Fig.~\ref{fig:21}, as has been recently done for the HW1 model~\cite{Hong:2009zw}. However, in this paper we want to be able to test the pion-pole approximation. In order to do so, one should trade the $\pi^0\gamma\gamma$ form factor for the more general objects $P^0\gamma\gamma$ or $A^0\gamma\gamma$. The issue of on-shellness versus off-shellness of the pion has attracted some attention in recent years~\cite{Nyffeler:2009tw,Jegerlehner:2007xe,Dorokhov:2008pw}. Here we will follow the approach taken in~\cite{Nyffeler:2009tw} and characterize the degree of off-shellness entirely by the short-distance constraint 
\begin{eqnarray}
\lim_{Q^2\to\infty}F_{\pi^{0*}\gamma^*\gamma^*}(Q^2,Q^2,0)=-\frac{f_{\pi}}{3}\chi_0+\cdots~.
\label{shsh}
\end{eqnarray}
From the previous expression one readily sees that the degree of off-shellness is then regulated by the parameter $\chi_0$. Unfortunately, this parameter is only poorly known. We will discuss this issue and its impact on the $(g-2)_{\mu}$  at the end of this Section.

Our strategy to test the pion-pole approximation will be the following:
\begin{itemize}
\item [(i)] We will use the following ansatz for the form factor:
\begin{eqnarray}\label{intpol}
K(q_1^2,q_2^2)&=&1+\lambda\left(\frac{q_1^2}{q_1^2-m_{V}^2}+\frac{q_2^2}{q_2^2-m_{V}^2}\right)\nonumber\\
&+&\eta\frac{q_1^2 q_2^2}{(q_1^2-m_{V}^2)(q_2^2-m_{V}^2)}~,
\end{eqnarray}
first introduced in the context of $K_L\to \mu^+\mu^-$
decays~\cite{D'Ambrosio:1997jp}. This parametrization was
originally put forward to study the low-energy slope of three-point functions, and has been used in experimental studies of
$K_L\to \mu^+\mu^-$ and $K_L\to \pi^0 e^+e^-$.
 \item [(ii)] We will
promote  the ansatz in (\ref{intpol}) to an analytic
interpolator valid for any value of the photon momenta. Then,
using Eqs. \eqref{FandK} and \eqref{shsh}, we get an expression
for $\chi_0$ in terms of the linear slope $\lambda$
\begin{equation}\label{chiform}
 \chi_0=\displaystyle{\frac{N_c}{4\pi^2 f_\pi^2}}(1 +\lambda)~.
\end{equation}
\end{itemize}

Aside from being simple and with a very intuitive low-energy behavior, the parametrization of Eq.~(\ref{intpol}) has additional advantages. One of the most important is that the computation of $(g-2)_{\mu}$ turns out to be greatly simplified. In Ref.~\cite{Knecht:2001qf} it was shown that if the interpolator can be cast in the generic form:
\begin{eqnarray}\label{param}
F_{\pi^0\gamma^*\gamma^*}(q_1^2,q_2^2)&=&-\frac{N_c}{12\pi^2f_{\pi}}\Big[f(q_1^2)\nonumber\\
&-&\sum_i\frac{1}{q_2^2-m_i^2}g_i(q_1^2)\Big]~,
\end{eqnarray}
then, in the general two-loop expression for the pion contribution
to the HLBL scattering, all angular integrations can be performed
using the hyperspherical approach. The diagram of
Fig.~\ref{fig:21} can then be evaluated through the following
expression:
\begin{equation}
a_{\mu}^{\pi^0}=\left(\frac{\alpha_{em}}{\pi}\right)^3\Big\{a_{\mu(1)}^{\pi^0}+a_{\mu(2)}^{\pi^0}\Big\}~,
\end{equation}
where
\begin{eqnarray}
a_{\mu(1)}^{\pi^0}&=&\int_0^{\infty}dQ_1\int_0^{\infty}dQ_2\Big[w_1(Q_1,Q_2)G_1(Q_1,Q_2)\nonumber\\
&&+w_2(m_V,Q_1,Q_2)G_2(Q_1,Q_2)\Big]\label{a1}
\end{eqnarray}
and
\begin{eqnarray}
a_{\mu(2)}^{\pi^0}&=&\int_0^{\infty}dQ_1\int_0^{\infty}dQ_2 \Big[w_3(m_V,Q_1,Q_2)G_3(Q_1,Q_2)\nonumber\\
&&+w_3(m_{\pi},Q_1,Q_2)G_4(Q_1,Q_2)\Big]~.\label{a2}
\end{eqnarray}
In the previous expressions, $G_i$ are generalized form factors and $w_i$ weight factors, whose expressions are given in Appendix~A.

In other words, the contribution to the $(g-2)_{\mu}$ can be reduced to a more tractable double integral. It is not difficult to verify that the interpolator of Eq.~(\ref{intpol}) is of the form (\ref{param}) with
\begin{eqnarray}
f(q^2)&=&1+\lambda+(\lambda+\eta)\frac{q^2}{q^2-m_V^2}~,\\
g_V(q^2)&=&-m_V^2\left[\lambda+\eta\frac{q^2}{q^2-m_V^2}\right]~.
\end{eqnarray}

Notice that the interpolator of Eq.~(\ref{intpol}) is assumed not to depend on the pion momentum. This simplifying assumption, together with the specific form of Eq.~(\ref{param}), is crucial to be able to use Eqs.~(\ref{a1}) and (\ref{a2}).

Our parametrization has 3 free parameters, $\lambda$, $\eta$ and $m_V$ which will be determined by combining the following constraints:
\begin{eqnarray}\label{system}
\frac{\lambda}{m_V^2}&=&(-1.76\pm0.22)\,\, {\mathrm{GeV}}^{-2}~,\label{expalpha}\\
1+2\lambda+\eta&=&0~,\label{1plus2alphaplusbeta}\\
\lambda+\eta&=&-\frac{4\pi^2f_{\pi}^2}{3m_V^2}~.
\end{eqnarray}
The first one is a low-energy experimental constraint on the slope of the pion form factor (the $\hat\alpha$ parameter of the previous Section), while the last two are short-distance information, namely the requirement that $K(Q^2,Q^2)$ at high energies does not go like a constant (first equation) and that its $1/Q^2$ behavior has the right coefficient (see Eq.~(\ref{OPEQ}), with $N_c=3$).

\begin{table*}[t]
\begin{tabular}{cccccc}
\hline
Model &\,\,\,\,\,\,\,\,\,\, $w_1G_1$ &\,\,\,\,\,\,\,\,\,\, $w_2G_2$&\,\,\,\,\,\,\,\,\,\, $w_3G_3$&\,\,\,\,\,\,\,\,\,\, $w_4G_4$&\,\,\,\,\,\,\,\,\,\,$a_{\mu}$\\
\hline
LMD &\,\,\,\,\,\,\,\,\,\, $+0.015$ &\,\,\,\,\,\,\,\,\,\, $+0.042$ &\,\,\,\,\,\,\,\,\,\, $+0.0016$ &\,\,\,\,\,\,\,\,\,\,$-0.0002$&\,\,\,\,\,\,\,\,\,\,$7.3\cdot 10^{-10}$ \\
DIP$_{\hat{\alpha}}$ &\,\,\,\,\,\,\,\,\,\, $+0.018(3)$ &\,\,\,\,\,\,\,\,\,\, $+0.034(4)$ &\,\,\,\,\,\,\,\,\,\, $+0.0016$ &\,\,\,\,\,\,\,\,\,\,$-0.0002$&\,\,\,\,\,\,\,\,\,\,$6.7(3)\cdot 10^{-10}$ \\
DIP$_{m_{\rho}}$ &\,\,\,\,\,\,\,\,\,\, $+0.015$ &\,\,\,\,\,\,\,\,\,\, $+0.043$ &\,\,\,\,\,\,\,\,\,\, $+0.0016$ &\,\,\,\,\,\,\,\,\,\,$-0.0002$&\,\,\,\,\,\,\,\,\,\,$7.35\cdot 10^{-10}$ \\
\hline
\end{tabular}
{\caption{Determination of $a_{\mu}$ with the DIP parametrization: in the second row, the mass scale is determined dynamically while, in the third row, $m_V=m_{\rho}$. Comparison is made with the LMD model. We explicitly show the contribution from the different generalized form factors $G_i$.}}\label{tab2}
\end{table*}

Using as input parameters
\begin{eqnarray}
m_{\mu}&=&105.658367(4)\,{\mathrm{MeV}}~,\nonumber\\
m_{\pi}&=&134.9766(6)\,{\mathrm{MeV}}~,\nonumber\\
f_{\pi}&=&92.4\,{\mathrm{MeV}}~,\nonumber\\
\alpha_{em}&=&1/137.03599976~,
\end{eqnarray}
the parameters of the model become
\begin{eqnarray}\label{resu}
\lambda&=&-0.73\pm0.05~,\nonumber\\
\eta&=&0.46^{+0.10}_{-0.13}~,\nonumber\\
m_V&=&(0.64^{+0.07}_{-0.06})\, {\mathrm{GeV}}~, \label{eq:lambdaeta}\nonumber\\
\chi_0&=&(2.42\pm0.17)\, {\mathrm{GeV}}^{-2}~,
\end{eqnarray}
where the uncertainties are due to the experimental error on the slope. 

Several comments are relevant at this point:
\begin{itemize}
\item[(i)] The values of  $\hat{\beta}$ from the HW1 and HW2 models in
Table~I are in very good agreement with the value of $\eta~(\sim
\hat{\beta} m_V^4)$ in Eqs.~\eqref{eq:lambdaeta}. We have
already mentioned that also the linear slope, $\lambda$, is very
well reproduced by HW1 and HW2 in Table~I. 

\item[(ii)] The value we find for the mass scale has to be interpreted as an effective measure of the relevant scale for the problem. Therefore, we confirm that $m_{\rho}$ is indeed very close to the natural scale to estimate the pion exchange contribution.

\item[(iii)] The
prediction for $ \chi_0$ in \eqref{eq:lambdaeta} is obtained from
the requirement that DIP is a good interpolator also for an off-shell pion and uses only the phenomenological linear
slope, $\lambda$ (and not the holographic one). However, the result would
have been very similar by trading the phenomenological linear
slope with the holographic values predicted by HW1 and HW2. 

\item[(iv)]
In order to test the stability and improve the accuracy of the results on HLBL one can add an extra pole to the interpolator. We shall follow this
line of thought later on.
\end{itemize}

With the values given in Eq.~(\ref{resu}), the prediction for the anomalous magnetic moment is
\begin{equation}
a_{\mu}^{\pi^0}=6.7(3)\cdot 10^{-10}~.
\end{equation}
In Table~III the comparison is made with the LMD result quoted in Ref.~\cite{Knecht:2001qf}. In LMD models it is common to identify $m_V\equiv m_{\rho}$. In contrast, in our case the mass scale is determined dynamically by the constraints. The first thing to realize is that the contributions for each integral (each column in Table~III) are in good agreement, even though the interpolators are very different and $m_V=0.64\,{\mathrm{GeV}}\neq m_{\rho}$. However, if we redo our analysis but now set $m_V=m_{\rho}$ from the start (and ignore Eq.~(\ref{expalpha})), then there is {\emph{full}} agreement with LMD (third line of Table~III). Thus, the previous exercise implies that the difference between LMD and our approach is entirely due to the choice of mass scale, which in our case is determined self-consistently. Therefore, while lowest meson dominance might be a reasonable strategy when little information about the correlator is known, with the present status of the HLBL, it seems more justified to take full advantage of the available information.

\subsection{Extension of the DIP ansatz}
The interpolator considered in the previous Section provided an
estimate  for the HLBL that fulfilled the leading long and short-distance constraints from $F_{\pi^0\gamma^*\gamma^*}$. However, as we
stated in the introduction, in this paper we want to investigate the
dependence on the low-energy parameters $\hat\beta$ and $\hat\gamma$ and estimate the impact of
the pion away from its mass shell.

The parameters $\hat\beta$ and $\hat\gamma$ were
computed in Section~\ref{secIII}, Eqs.\eqref{alphaholography} and
\eqref{betaholography}. In order to understand their potential
importance for the HLBL, in Fig.~5 we show the shape of the weight
functions entering the dominant $a_{\mu(1)}^{\pi^0}$. As pointed
out in~\cite{Knecht:2001qf}, the relevant contributions are highly
peaked at low values of momenta, $0\leq Q^2\lesssim 0.5$ GeV.
Therefore, this suggests that more information on (very) low
energies can help improve the determination of $a_{\mu}$.

In order to check the stability of the results obtained so far,
in particular Eqs.~\eqref{eq:lambdaeta}, we consider a
generalization of the DIP interpolator by adding an extra pole:
\begin{eqnarray}\label{intpol2a}
K(q_1,q_2)=1+\sum_i^2\lambda_i\left(\frac{q_1^2}{q_1^2-m_{i}^2}+\frac{q_2^2}{q_2^2-m_{i}^2}\right)\nonumber\\
+\sum_i^2\eta_i\frac{q_1^2 q_2^2}{(q_1^2-m_{i}^2)(q_2^2-m_{i}^2)}~.\,\,\,\,\,
\end{eqnarray}
Notice that this increase in the number of poles, which is also natural from a large-$N_c$
perspective, makes Eq.~(\ref{intpol2a}) still compatible with the
general form of Eq.~(\ref{param}), now with
\begin{eqnarray}
f(q^2)&=&1+\sum_i^2\lambda_i+\sum_i^2(\lambda_i+\eta_i)\frac{q^2}{q^2-m_i^2}~,\,\,\,\,\,\,\,\\
g_i(q^2)&=&-m_i^2\left[\lambda_i+\eta_i\frac{q^2}{q^2-m_i^2}\right]~.
\end{eqnarray}
Therefore, the HLBL contribution can still be computed with the
double  integrals of Eqs.~(\ref{a1}) and~(\ref{a2}).
\begin{figure}[t]
\begin{center}
\includegraphics[width=6cm]{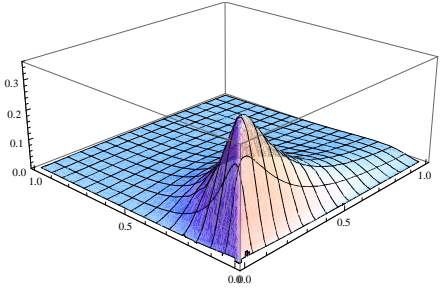}
\end{center}
\caption{The integrand of Eq.~(\ref{a1}).\label{fig:1} }
\end{figure}
By expanding Eq.~(\ref{intpol2a}) at low and high energies and
matching  it afterwards to the information coming from QCD short
distances and low-energy (experimental and holographic) input, the
constraints we would like to impose take the following form:
\begin{itemize}

\item[(a)] Long-distance constraints:
\begin{eqnarray}\label{system1}
\sum_i^2\frac{\lambda_i}{m_i^2}&=&(-1.76\pm0.22)\,\,{\mathrm{GeV}}^{-2}~,\,\,{\mathrm{(exp.)}}\label{first}\\
\sum_i^2\frac{\eta_i}{m_i^4}&=&(3.33\pm0.32)\,\,{\mathrm{GeV}}^{-4}~,\,\,{\mathrm{(pred.)}}\label{second}\\
\sum_i^2\frac{\lambda_i}{m_i^4}&=&(-2.84\pm0.21)\,\,{\mathrm{GeV}}^{-4}~,\,\,{\mathrm{(pred.)}}\label{third}
\end{eqnarray}
which correspond to the low-energy parameters ${\hat{\alpha}}$,
${\hat{\beta}}$  and ${\hat{\gamma}}$ defined in the previous
Section. The value for ${\hat{\alpha}}$ is taken from experiment,
while ${\hat{\beta}}$ and
${\hat{\gamma}}$ are estimated from the spread of values
predicted by the HW1, HW2 and SW models and given in Eqs.~\eqref{alphaholography}
and \eqref{betaholography}.

\item[(b)] Short-distance
constraints:
\begin{eqnarray}
1+2\sum_i^2\lambda_i+\sum_i^2\eta_i&=&0~,\label{firstsh}\\
\sum_i^2m_i^2(\lambda_i+\eta_i)&=&-\frac{4\pi^2f_{\pi}^2}{3}~,\label{ll}~,\\
1+\sum_i^2\lambda_i&=&\frac{4\pi^2f_{\pi}^2}{3}\chi_0~.\label{sho}
\end{eqnarray}
\end{itemize}
The first two are the ones already discussed in the previous Sections, namely the absence of constant terms and the matching
of the $1/Q^2$ coefficient. The last condition comes from Eq.~(\ref{shsh}).

We have already mentioned that the DIP interpolator (and its extensions) do not depend, by construction, on the pion momentum. While this allows us to use Eqs.~(\ref{a1}) and (\ref{a2}), we want to note at this point that this simplifying assumption implies that Eq.~(\ref{sho}) above and the Brodsky-Lepage constraint cannot be satisfied simultaneously (unless $\chi_0=0$). However, there are reasons to expect that
the Brodsky-Lepage constraint plays a minor role in HLBL. First and foremost,
the kernel entering $(g-2)_{\mu}$ is peaked at low energies (see
Fig.~\ref{fig:1}) and presumably information on the high-energy
region will have a negligible impact on the anomalous magnetic
moment. In this respect, it is quite reassuring that our analysis,
in which the $F_{\pi^0\gamma^*\gamma^*}$ low-energy region is
approximately captured by Eqs.~(\ref{first})-(\ref{third}), turns
out to be in very good agreement with analyses that comply with
the Brodsky-Lepage
constraint~\cite{Knecht:2001qf,Nyffeler:2009tw}.
 
The detailed solution of the system of
Eqs.~(\ref{first})-(\ref{sho}) is given in Appendix B. The main
conclusion of the analyses is that there is a remarkable stability
when either $\hat\beta$ or $\hat\gamma$ are added as input.
Furthermore, the uncertainty on $(g-2)_{\mu}$ is dominated by
$\hat\alpha$, while the effect from $\hat\beta$ and $\hat\gamma$
is subleading. Our estimate for the pion contribution to the HLBL
is
\begin{equation}\label{finalresult}
a_{\mu}=6.54(25)\cdot 10^{-10}~.
\end{equation}


\subsection{Impact of $\chi_0$ on $(g-2)_\mu$}

Since $\chi_0$ cannot be accessed experimentally, one has to
resort to  nonperturbative techniques to estimate it. However,
the different available estimates  show strong discrepancies:
while recent results based on QCD sum rules~\cite{Ball:2002ps} or
exclusive $B$ decays~\cite{Rohrwild:2007yt} seem to favor values
hovering around $\chi_0\sim 2-4$ GeV$^{-2}$, estimates based on
the axial anomaly~\cite{Vainshtein:2002nv,Gorsky:2009ma}, Pade
approximants~\cite{Cata:2009fd} or the original sum rule
determination of~\cite{Ioffe:1983ju} point at much higher values,
$\chi_0\sim 8-11$ GeV$^{-2}$. A recent result in the context of
holographic QCD~\cite{Cappiello:2010tu} also pointed out the
possibility that $\chi_0\sim 0$. Most of these estimates are given
without reference to the renormalization scale. However, its low-energy scale evolution was studied in~\cite{Cata:2007ns} and does
not account for the discrepancies found. Therefore, the best one
can do at present is to consider a conservative $0\leq \chi_0\leq
8.9$ GeV$^{-2}$. 

The result in Eq.~(\ref{finalresult}) is obtained assuming that $\chi_0\sim (1-3)$
GeV$^{-2}$, which seems to be the range of values favored by our
interpolator (see Table~V in Appendix B). However, larger values of $\chi_0\sim 9$ GeV$^{-2}$, which at present
cannot be excluded, might induce a shift of at most $15\%$. Given
the precision presently needed for the $(g-2)_{\mu}$, we believe
that a better understanding of $\chi_0$ is essential.


\section{Conclusions}\label{secV}
The muon anomalous magnetic moment is one of the most precisely
measured quantities  in particle physics, and thus a key parameter
to test the appearance of physics beyond the Standard Model. With
experimental accuracies expected to soon reach the $10^{-10}$
level, it is essential to have the theoretical uncertainties in
the hadronic light-by-light scattering contribution well under
control.

In this paper we have studied the electromagnetic pion form factor
$F_{\pi^0\gamma\gamma}$ with a set of holographic models. We have shown that only a restricted set of the holographic
models studied in Section~\ref{secII}, {\it i.e} HW1, HW2 and a modified
version of SW, are able to predict a value of the linear slope
compatible with the experimental data. Within this restricted set
of models we have then obtained, in Section~\ref{secIII}, predictions for the quadratic slopes
\begin{eqnarray}
{\hat{\beta}}=3.33(32)\,{\mathrm{GeV}}^{-4},\nonumber\\ 
{\hat{\gamma}}=2.84(21)\,{\mathrm{GeV}}^{-4}.
\end{eqnarray}
Notice that, because of the holographic models used, the previous predictions are sensitive to the way chiral symmetry breaking is implemented and whether the spectrum is Regge-like or not (SW versus HW). This sensitivity is contained in the quoted uncertainty.
 
We have then introduced an interpolator for the form factor that encodes both short distances (coming from QCD) and long distances (the quadratic slopes $\hat\beta$ and $\hat\gamma$ predicted above together with the experimental linear slope $\hat\alpha$) and computed the pion exchange HLBL contribution to the $(g-2)_{\mu}$. Following~\cite{Nyffeler:2009tw}, we have also estimated the effects due to a departure from the pion-pole approximation through a new short-distance constraint for $F_{\pi{^0}^{*}\gamma^*\gamma^*}$, which is controlled by the parameter $\chi_0$. Our final number for the (pion exchange) HLBL contribution is
\begin{equation}\label{last}
a_{\mu}^{\pi^0}=(6.54\pm0.25)\cdot 10^{-10}~,
\end{equation}
which depends on the slope ($\hat\alpha$) and curvature ($\hat\beta$ and $\hat\gamma$) of $F_{\pi^0\gamma^*\gamma^*}$ as well as on the parameter $\chi_0$. Our analysis shows that, while $\hat\beta$ and $\hat\gamma$ are important ingredients, the uncertainty is currently dominated by the experimental accuracy on $\hat\alpha$. In view of this, improvements on the slope $\hat\alpha$ should be a priority. Actually, this is one of the main points in the scientific program for the KLOE-2 proposal~\cite{Babusci:2010ym} at DA$\Phi$NE in Frascati. With high enough statistics, it is even feasible that the holographic predictions for $\hat\beta$ and ${\hat{\gamma}}$ could be tested.

Another sizable source of uncertainty comes from the parameter $\chi_0$. Our result in Eq.~(\ref{last}) assumes $0\lesssim\chi_0\lesssim 3$ GeV$^{-2}$. While numbers hovering around $\chi_0\sim 2$ GeV$^{-2}$ seem to be favored by our analysis and also by different theoretical estimations, higher values ($\chi_0\sim 9$ GeV$^{-2}$) have also been reported in the literature. These higher values for $\chi_0$ would induce a $(10-15)\%$ shift on the HLBL contribution. Therefore, we want to emphasize the need to put stronger bounds on the current value of $\chi_0$. In particular, a lattice determination of $\chi_0$ would be extremely useful.

Finally, the effects of model dependence in our approach are evaluated in two ways: first, through the spread of values for the low-energy quartic terms in $F_{\pi^0\gamma^*\gamma^*}$, coming from the different holographic models; and second, using different versions for the parametrization of $F_{\pi^0\gamma^*\gamma^*}$. The resulting uncertainties are negligible, as it can be seen in Table~IV.

We have also compared our analysis with LMD
models. In Section~\ref{secIII} we have investigated the issue of
LMD in $F_{\pi^0\gamma^*\gamma^*}$ with holographic models, which
contain a full spectrum of resonances. Our conclusion is that LMD
is a sound approximation. This is confirmed later on in our
evaluation of the HLBL: the mass scale of our interpolator, which
is determined dynamically, is always close to $m_{\rho}$. In other
words, the LMD observed in $(g-2)_{\mu}$ is a consequence of LMD
in the $F_{\pi^0\gamma^*\gamma^*}$, a feature that holographic
realizations successfully predict. A natural consequence of LMD is that the curvature $\hat{\gamma}$
of the pion form factor can be estimated in terms of the slope
$\hat{\alpha}$
\begin{equation}
\hat\gamma=-\frac{\hat\alpha}{m_V^2}\sim 2.97\,\, {\mathrm{GeV}}^{-4}~,
\end{equation}
which agrees well with the holographic prediction. Also
$\hat\beta$ can be estimated from LMD, but an additional
ingredient has to be provided.  If one uses the first {\emph{short-distance}} constraint coming from the DIP interpolator and
combines it with lowest vector dominance setting $m_V\sim
m_{\rho}$, one finds
\begin{equation}
\hat\beta=-\frac{2m_V^2\hat\alpha+1}{m_V^4}\sim 3.10\,\,{\mathrm{GeV}}^{-4}~,
\end{equation}
again in remarkable agreement with the holographic prediction.
This also means that the {\emph{short-distance}} constraint in the
DIP interpolators is satisfied to a very good approximation by the
low-energy values predicted by holographic models. This
compatibility between long and short distances is of considerable
importance. Since the previous results are mostly based on
$\rho$-dominance in conjunction with our DIP interpolator, we
conjecture that they might have a broader applicability, not just
in $\pi^0\to\gamma\gamma$ but also in leptonic and semileptonic
decays, {\it{e.g.}}, $K_L\to\mu^+\mu^-$ and $K_L\to\pi^0e^+e^-$.
In particular, it opens up a way to estimate $\hat\beta$ in
$K_L\to (\mu^+\mu^-, \pi^0e^+e^-)$.

However, when it comes to the evaluation of the HLBL,  within the
current precision LMD is probably not good enough. We have shown
that the discrepancy between our result and the one from LMD
models originates entirely from the different choices of the
effective mass scale: while LMD assumes $m_V=m_{\rho}$, in our
approach the mass is determined from the matching conditions.
Since this deviation is comparable to the typical uncertainty on
HLBL, assuming $m_V=m_{\rho}$ does not seem justified.


\section*{Acknowledgments}
We want to thank A.~Nyffeler for correspondence and useful comments on the first version of this manuscript. O.~C.~wants to thank the University of Naples for very pleasant stays during the different stages of this work. L.~C.~ and G.~D'A.~ are supported in part by the EU under Contract No. MTRN-CT-2006-035482 (FLAVIAnet), by MIUR, Italy, under Project No. 2005-023102 and by Fondo dipartimentale per la ricerca 2009. O.~C.~is supported by the EU under Contract No. MTRN-CT-2006-035482 (FLAVIAnet) and by MICINN, Spain under Grants No. FPA2007-60323
and Consolider-Ingenio 2010 CSD2007-00042 –CPAN–.\\

\appendix
\section{}
The generalized form factors $G_i$ entering Eqs.~(\ref{a1},\ref{a2}) are given by
\begin{eqnarray}
G_1(x,y)&=&-\frac{N_c}{12\pi^2f_{\pi}}f(-x^2)F_{\pi^0\gamma^*\gamma^*}(-y^2,0)~,\nonumber\\
G_2(x,y)&=&-\frac{N_c}{12\pi^2f_{\pi}}\frac{g(-x^2)}{m_V^2}F_{\pi^0\gamma^*\gamma^*}(-y^2,0)~,\nonumber\\
G_3(x,y)&=&-\frac{N_c}{12\pi^2f_{\pi}}\frac{g(0)}{m_{\pi}^2-m_V^2}F_{\pi^0\gamma^*\gamma^*}(-x^2,-y^2)~,\nonumber\\
G_4(x,y)&=&-\frac{N_c}{12\pi^2f_{\pi}}f(0)F_{\pi^0\gamma^*\gamma^*}(-x^2,-y^2)-G_3(x,y)~,\nonumber\\
\end{eqnarray}
while the weight factors $w_i$ take the form
\begin{widetext}
\begin{eqnarray}
w_1(x,y)&=&\frac{\pi^2}{6m_{\mu}^2xy(y^2+m_{\pi}^2)}\left[4(y^2-2m_{\mu}^2)(x^2-y^2)^2 \log(1+\lambda(0,x,y))\right.\nonumber\\
&+&\left.\big((x^6+y^6)-x^2y^4-3x^4y^2+\frac{x^4y^6}{2m_{\mu}^4}-\frac{x^4y^4}{m_{\mu}^2}-4m_{\mu}^2x^2y^2-(x^2-y^2)^2\eta_0\big)\left(1-\frac{\xi(x)}{x^2}\right)\right.\nonumber\\
&+&\left.y^2(y^4-4m_{\mu}^4)\frac{\xi(x)}{m_{\mu}^2}-x^4(y^2-2m_{\mu}^2)^2\frac{\xi(y)}{2m_{\mu}^4}+(y^2-2m_{\mu}^2)(x^2y^2-2m_{\mu}^2(x^2+y^2))\frac{\xi(x)\xi(y)}{2m_{\mu}^4}\right]~,\,\,\,\,\,\,\,\,\,\,\,\,\,\,
\end{eqnarray}
\begin{eqnarray}
w_2(m,x,y)&=&\frac{\pi^2}{6m_{\mu}^2xy(y^2+m_{\pi}^2)}\left[4(y^2-2m_{\mu}^2)(x^2-y^2)^2 \log(1+\lambda(0,x,y))\right.\nonumber\\
&-&\left.4(y^2-2m_{\mu}^2)(m^4+(x^2-y^2)^2+2m^2(x^2+y^2))\log(1+\lambda(m,x,y))\right.\nonumber\\
&+&\left.\left\{(m^4+2m^2(x^2+y^2)+(x^4+y^4)-2x^2y^2)\eta_m-\frac{m^2}{m_{\mu}^2}x^2(y^2-2m_{\mu}^2)\xi(y)+\frac{m^2}{m_{\mu}^2}x^2y^4\right.\right.\nonumber\\
&-&\left.\left.(x^2-y^2)^2\eta_0-m^6-3m^4(x^2+y^2)-3m^2(x^4+y^4)-2m^2x^2y^2\right\}\left(1-\frac{\xi(x)}{x^2}\right)\right]~,
\end{eqnarray}
\begin{eqnarray}
w_3(m,x,y)&=&\frac{\pi^2}{6m_{\mu}^2xy}\left[4(x^2-y^2)(m_{\mu}^2(y^2-x^2)+2x^2y^2)\log(1+\lambda(0,x,y))+4\log(1+\lambda(m,x,y))\right.\nonumber\\
&\times&\left.(m^4m_{\mu}^2+(x^2-y^2)(m_{\mu}^2(x^2-y^2)-2x^2y^2)+2m^2(m_{\mu}^2(x^2+y^2)+x^2y^2))+m^4(x^2+y^2)\right.\nonumber\\
&+&\left.2m^2(x^4+y^4)-m^2x^2y^2-m^2(m^2+2x^2+y^2)\xi(x)-m^2(m^2-3x^2+2y^2)\xi(y)-m^2\xi(x)\xi(y)\right.\nonumber\\
&\!\!\!\!\!\!\!\!\!\!\!\!\!\!\!\!\!\!\!\!+&\!\!\!\!\!\!\!\!\!\!\!\!\left.\left\{(x^2+y^2)\eta_0-(m^2+x^2+y^2)\eta_m\right\}(x^2-\xi(x))+\left\{(y^2-3x^2)\eta_0-(m^2+y^2-3x^2)\eta_m\right\}(y^2-\xi(y))\right]~.\nonumber\\
\end{eqnarray}
\end{widetext}
In the previous expressions we have defined
\begin{equation}
\lambda(m,x,y)=\frac{(m^2\!+\!x^2\!+\!y^2\!-\eta_m)(x^2-\xi(x))(y^2-\xi(y))}{8m_{\mu}^2x^2y^2}~,
\end{equation}
\begin{equation}
\eta_m=\sqrt{(m^2+x^2+y^2)^2-4x^2y^2}~,
\end{equation}
and
\begin{eqnarray}
\xi(z)&=&\sqrt{z^4+4m_{\mu}^2z^2}~.
\end{eqnarray}


\section{}
We start from the generalized interpolator:
\begin{eqnarray}\label{intpol2}
K(q_1,q_2)=1+\sum_i^2\lambda_i\left(\frac{q_1^2}{q_1^2-m_{i}^2}+\frac{q_2^2}{q_2^2-m_{i}^2}\right)\nonumber\\
+\sum_i^2\eta_i\frac{q_1^2 q_2^2}{(q_1^2-m_{i}^2)(q_2^2-m_{i}^2)}~,\,\,\,\,\,
\end{eqnarray}
whose expansion at low and high energies leads to the system of constraints of Eqs.~(\ref{first})-(\ref{sho}). Our purpose is to investigate the impact of the low-energy parameters $\hat\beta$, $\hat\gamma$ and $\chi_0$. This we will do by varying the combination of constraints to be imposed on Eq.~(\ref{intpol2}).

First let us concentrate on the parameters $\hat\beta$. The system of equations we have to solve is then
\begin{eqnarray}\label{eqsbeta}
\sum_i^2\frac{\lambda_i}{m_i^2}&=&-1.76(22)\,\,{\mathrm{GeV}}^{-2}~,\,\,{\mathrm{(exp.)}}\nonumber\\
\sum_i^2\frac{\eta_i}{m_i^4}&=&3.33(32)\,\,{\mathrm{GeV}}^{-4}~,\,\,{\mathrm{(pred.)}}\nonumber\\
1+2\sum_i^2\lambda_i+\sum_i^2\eta_i&=&0~,\nonumber\\
\sum_i^2m_i^2(\lambda_i+\eta_i)&=&-\frac{4\pi^2f_{\pi}^2}{3}~.
\end{eqnarray}
In order to ensure a solution, we will set $m_2=m_{\rho}=0.775$ GeV and $\lambda_2=0$. The interpolator, therefore, reduces to
\begin{eqnarray}
K_{(1)}(q_1,q_2)=1+\lambda\left(\frac{q_1^2}{q_1^2-m_{1}^2}+\frac{q_2^2}{q_2^2-m_{1}^2}\right)\nonumber\\
+\sum_i^2\eta_i\frac{q_1^2 q_2^2}{(q_1^2-m_{i}^2)(q_2^2-m_{i}^2)}~.
\end{eqnarray}
In order to check for stability, we will also consider the setting in which $m_2=m_{\rho}=0.775$ GeV and $\eta_2=0$. The corresponding interpolator then reads
\begin{eqnarray}
K_{(2)}(q_1,q_2)=1+\sum_i^2\lambda_i\left(\frac{q_1^2}{q_1^2-m_{i}^2}+\frac{q_2^2}{q_2^2-m_{i}^2}\right)\nonumber\\
&&\!\!\!\!\!\!\!\!\!\!\!\!\!\!\!\!\!\!\!\!\!\!\!\!\!\!\!\!\!\!\!\!\!\!\!\!\!\!\!\!\!\!\!\!\!\!\!\!\!\!\!\!\!\!\!\!\!\!\!\!\!\!\!\!\!\!\!\!\!\!\!\!\!\!\!\!\!+\eta\frac{q_1^2 q_2^2}{(q_1^2-m_{1}^2)(q_2^2-m_{1}^2)}~.
\end{eqnarray}
The advantage of dealing with two generalized expressions is that we will be able to assess the dependence of our results on the specific form of the interpolator. We will denote the previous setting as DIP$^{(i)}_{\hat\beta;m_2=m_{\rho}}$, where the superscript refers to the different interpolator used ($K_{(1)}$ or $K_{(2)}$) and the subscript emphasizes that $m_2$ has been fixed and that $\hat\beta$ has been used as input. The results for the HLBL are shown in the second and third rows of Table~IV and can be compared with the result from DIP$_{\hat\alpha}$, which corresponds to the single resonance analysis (see Table~III) in the main text. The stability of the results is remarkable, as can be seen from the fact that the dependence on the parametrization ($K_{(1)}$ or $K_{(2)}$) turns out to be very mild.

An analogous analysis can be done for $\hat\gamma$. The only difference is that now the system of matching equations looks like
\begin{eqnarray}\label{eqsgamma}
\sum_i^2\frac{\lambda_i}{m_i^2}&=&-1.76(22)\,\,{\mathrm{GeV}}^{-2}~,\,\,{\mathrm{(exp.)}}\nonumber\\
\sum_i^2\frac{\lambda_i}{m_i^4}&=&-2.84(21)\,\,{\mathrm{GeV}}^{-4}~,\,\,{\mathrm{(pred.)}}\nonumber\\
1+2\sum_i^2\lambda_i+\sum_i^2\eta_i&=&0~,\nonumber\\
\sum_i^2m_i^2(\lambda_i+\eta_i)&=&-\frac{4\pi^2f_{\pi}^2}{3}~.
\end{eqnarray}
The entries in Table~IV are DIP$^{(i)}_{\hat\gamma;m_2=m_{\rho}}$. From the results, we see again that the different determinations agree within errors, but somehow the uncertainty associated with $\hat\alpha$ has grown.

A conservative estimate, taking into account the first five rows of Table~IV would give
\begin{equation}
a_{\mu}=6.4(5)\cdot 10^{-10}~.
\end{equation}
The main source of uncertainty comes from the slope ${\hat{\alpha}}$, while the effect of ${\hat{\beta}}$ or ${\hat{\gamma}}$ is subleading. Thus, an improvement of the CELLO data on the slope of the $\pi^0\gamma\gamma$ form factor can help decrease the theoretical uncertainty on the HLBL.

The previous estimate for $a_{\mu}$ most probably overestimates the uncertainty. Notice that the results for DIP$_{\hat{\gamma}}^{(1,2)}$ give lower values of $a_{\mu}$ (due to a strong suppression of $w_1G_1$) and with bigger uncertainties from $\hat\alpha$. This might be a consequence of the strong correlation between ${\hat{\alpha}}$ and ${\hat{\gamma}}$ in the interpolator we are using and, therefore, such an uncertainty might be misleading. Relying on DIP$_{\hat{\beta}}^{(1,2)}$ and DIP$_{\hat\alpha}$ only, a more realistic estimate is
\begin{equation}
a_{\mu}=6.54(25)\cdot 10^{-10}~.
\end{equation}

\begin{table*}[t]
\begin{tabular}{cccccc}
\hline
Model &\,\,\,\,\,\,\,\,\,\, $w_1G_1$ &\,\,\,\,\,\,\,\,\,\, $w_2G_2$&\,\,\,\,\,\,\,\,\,\, $w_3G_3$&\,\,\,\,\,\,\,\,\,\, $w_4G_4$&\,\,\,\,\,\,\,\,\,\,$a_{\mu}$\\
\hline
DIP$_{\hat{\alpha}}$ &\,\,\,\,\,\,\,\,\,\, $+0.018(3)$ &\,\,\,\,\,\,\,\,\,\, $+0.034(4)$ &\,\,\,\,\,\,\,\,\,\, $+0.0016$ &\,\,\,\,\,\,\,\,\,\,$-0.0002$&\,\,\,\,\,\,\,\,\,\,$6.7(3)\cdot 10^{-10}$
\\
DIP$^{(1)}_{{\hat{\beta}};m_2=m_{\rho}}$ &\,\,\,\,\,\,\,\,\,\, $+0.014$ &\,\,\,\,\,\,\,\,\,\, $+0.037$ &\,\,\,\,\,\,\,\,\,\, $+0.0015$ &\,\,\,\,\,\,\,\,\,\,$-0.0002$&\,\,\,\,\,\,\,\,\,\,$6.52(15)(10)\cdot 10^{-10}$ \\
DIP$^{(2)}_{{\hat{\beta}};m_2=m_{\rho}}$ &\,\,\,\,\,\,\,\,\,\, $+0.014$ &\,\,\,\,\,\,\,\,\,\, $+0.037$ &\,\,\,\,\,\,\,\,\,\, $+0.0015$ &\,\,\,\,\,\,\,\,\,\,$-0.0002$&\,\,\,\,\,\,\,\,\,\,$6.55(21)(6)\cdot 10^{-10}$ \\
DIP$^{(1)}_{{\hat{\gamma}};m_2=m_{\rho}}$ &\,\,\,\,\,\,\,\,\,\, $+0.004$ &\,\,\,\,\,\,\,\,\,\, $+0.047$ &\,\,\,\,\,\,\,\,\,\, $+0.0015$ &\,\,\,\,\,\,\,\,\,\,$-0.0002$&\,\,\,\,\,\,\,\,\,\,$6.09(81)(9)\cdot 10^{-10}$ \\
DIP$^{(2)}_{{\hat{\gamma}};m_2=m_{\rho}}$ &\,\,\,\,\,\,\,\,\,\, $+0.002$ &\,\,\,\,\,\,\,\,\,\, $+0.047$ &\,\,\,\,\,\,\,\,\,\, $+0.0015$ &\,\,\,\,\,\,\,\,\,\,$-0.0002$&\,\,\,\,\,\,\,\,\,\,$6.21(77)(7)\cdot 10^{-10}$ \\
\hline
DIP$^{(1)}_{{\hat{\beta}};0<\chi_0<8.9}$ &\,\,\,\,\,\,\,\,\,\, $[+0.003;+0.047]$ &\,\,\,\,\,\,\,\,\,\, $[+0.043;+0.022]$ &\,\,\,\,\,\,\,\,\,\, $[+0.0015;+0.0016]$ &\,\,\,\,\,\,\,\,\,\,$-0.0002$&\,\,\,\,\,\,\,\,\,\,$[5.9;8.9]\cdot 10^{-10}$ \\
DIP$^{(2)}_{{\hat{\beta}};0<\chi_0<4.4}$ &\,\,\,\,\,\,\,\,\,\, $[+0.002;+0.027]$ &\,\,\,\,\,\,\,\,\,\, $[+0.044;+0.025]$ &\,\,\,\,\,\,\,\,\,\, $+0.0015$ &\,\,\,\,\,\,\,\,\,\,$-0.0002$&\,\,\,\,\,\,\,\,\,\,$[6.0;6.7]\cdot 10^{-10}$ \\
DIP$^{(2)}_{{\hat{\gamma}};0<\chi_0<6.4}$ &\,\,\,\,\,\,\,\,\,\, $[+0.006;+0.035]$ &\,\,\,\,\,\,\,\,\,\, $[+0.043;+0.022]$ &\,\,\,\,\,\,\,\,\,\, $[+0.0015;+0.0016]$ &\,\,\,\,\,\,\,\,\,\,$-0.0002$&\,\,\,\,\,\,\,\,\,\,$[6.3;7.3]\cdot 10^{-10}$ \\
\hline
\end{tabular}
{\caption{Determination of $a_{\mu}$ with the generalized DIP parametrizations and their comparison with the DIP$_{\hat\alpha}$ model. We explicitly show the contribution from the different generalized form factors $G_i$. The errors in parentheses are the ones induced by ${\hat{\alpha}}$ followed by ${\hat{\beta}}$ or ${\hat{\gamma}}$. For the last 3 rows, where $\chi_0$ is inside a range, we show the values for the endpoints.}}\label{tab3}
\end{table*}
\begin{table*}[t]
\begin{tabular}{ccccc}
\hline
Model &\,\,\,\,\,\,\,\,\,\, ${\hat{\alpha}}$ (GeV$^{-2}$) &\,\,\,\,\,\,\,\,\,\, ${\hat{\beta}}$ (GeV$^{-4}$)&\,\,\,\,\,\,\,\,\,\, ${\hat{\gamma}}$ (GeV$^{-4}$)&\,\,\,\,\,\,\,\,\,\, $\chi_0$ (GeV$^{-2}$)\\
\hline
DIP$_{\hat{\alpha}}$ &\,\,\,\,\,\,\,\,\,\, $-1.76^*$ &\,\,\,\,\,\,\,\,\,\, $2.67$ &\,\,\,\,\,\,\,\,\,\, $4.25$ &\,\,\,\,\,\,\,\,\,\,$2.42$\\
DIP$_{m_{\rho}}$ &\,\,\,\,\,\,\,\,\,\, $-1.35$ &\,\,\,\,\,\,\,\,\,\, $1.73$ &\,\,\,\,\,\,\,\,\,\, $2.25$ &\,\,\,\,\,\,\,\,\,\,$1.66$\\
DIP$^{(1)}_{\hat{\beta};m_2=m_{\rho}}$ &\,\,\,\,\,\,\,\,\,\, $-1.76^*$ &\,\,\,\,\,\,\,\,\,\, $3.33^*$ &\,\,\,\,\,\,\,\,\,\, $3.78$ &\,\,\,\,\,\,\,\,\,\,$1.61$\\
DIP$^{(2)}_{\hat{\beta};m_2=m_{\rho}}$ &\,\,\,\,\,\,\,\,\,\, $-1.76^*$ &\,\,\,\,\,\,\,\,\,\, $3.33^*$ &\,\,\,\,\,\,\,\,\,\, $3.88$ &\,\,\,\,\,\,\,\,\,\,$1.69$\\
DIP$^{(1)}_{\hat{\gamma};m_2=m_{\rho}}$ &\,\,\,\,\,\,\,\,\,\, $-1.76^*$ &\,\,\,\,\,\,\,\,\,\, $4.56$ &\,\,\,\,\,\,\,\,\,\, $2.84^*$ &\,\,\,\,\,\,\,\,\,\,$-0.81$\\
DIP$^{(2)}_{\hat{\gamma};m_2=m_{\rho}}$ &\,\,\,\,\,\,\,\,\,\, $-1.76^*$ &\,\,\,\,\,\,\,\,\,\, $5.23$ &\,\,\,\,\,\,\,\,\,\, $2.84^*$ &\,\,\,\,\,\,\,\,\,\,$-0.74$\\
\hline
DIP$^{(1)}_{{\hat{\beta}},0<\chi_0<8.9}$ &\,\,\,\,\,\,\,\,\,\, $-1.76^*$ &\,\,\,\,\,\,\,\,\,\, $3.33^*$ &\,\,\,\,\,\,\,\,\,\, $[3.10;-5\cdot10^{5}]$ &\,\,\,\,\,\,\,\,\,\,$[0;8.9]^*$\\
DIP$^{(2)}_{{\hat{\beta}},0<\chi_0<4.4}$ &\,\,\,\,\,\,\,\,\,\, $-1.76^*$ &\,\,\,\,\,\,\,\,\,\, $3.33^*$ &\,\,\,\,\,\,\,\,\,\, $[3.19;-3.18]$ &\,\,\,\,\,\,\,\,\,\,$[0;4.4]^*$\\
DIP$^{(2)}_{{\hat{\gamma}},0<\chi_0<6.4}$ &\,\,\,\,\,\,\,\,\,\, $-1.76^*$ &\,\,\,\,\,\,\,\,\,\, $[5.47,-18]$ &\,\,\,\,\,\,\,\,\,\, $2.84^*$ &\,\,\,\,\,\,\,\,\,\,$[0;6.4]^*$\\
\hline
\end{tabular}
{\caption{Predicted low-energy parameters from the different parametrizations. Asterisked quantities are input of the different DIP interpolators. For the last 3 rows, where $\chi_0$ is inside a range, we show the values for the endpoints. Notice from the last three lines that large values for $\chi_0$ are clearly disfavored.}}\label{tab4}
\end{table*}

Next we turn to the evaluation of the impact of $\chi_0$. For this we will keep $K_{(1)}$ and $K_{(2)}$ but, instead of setting $m_2=m_{\rho}$, we will add the short-distance constraint involving $\chi_0$ to Eq.~(\ref{eqsbeta}):
\begin{eqnarray}
\sum_i^2\frac{\lambda_i}{m_i^2}&=&-1.76(22)\,\,{\mathrm{GeV}}^{-2}~,\,\,{\mathrm{(exp.)}}\nonumber\\
\sum_i^2\frac{\eta_i}{m_i^4}&=&3.33(32)\,\,{\mathrm{GeV}}^{-4}~,\,\,{\mathrm{(pred.)}}\nonumber\\
1+2\sum_i^2\lambda_i+\sum_i^2\eta_i&=&0~,\nonumber\\
\sum_i^2m_i^2(\lambda_i+\eta_i)&=&-\frac{4\pi^2f_{\pi}^2}{3}~,\nonumber\\
1+\sum_i^2\lambda_i&=&\frac{4\pi^2f_{\pi}^2}{3}\chi_0~.
\end{eqnarray}
We will denote the resulting interpolators as DIP$^{(i)}_{\hat\beta;\chi_0}$. Accordingly, the addition of the $\chi_0$ constraint to Eqs.~(\ref{eqsgamma}) will be denoted as DIP$^{(i)}_{\hat\gamma;\chi_0}$. It is by no means trivial that those new systems of constraints will have a solution at all. It crucially depends on the interpolator. In this respect, notice that the DIP ansatz has the right kinematics to comply with Eq.~(\ref{shsh}), which is highly nontrivial.

The results are shown in the last three lines of Table~IV, where we scanned the solution over a wide interval for $\chi_0$, mostly covering all the values quoted in the literature so far. The range for $\chi_0$ is indicated in each case. While, in principle, we took $0\leq \chi_0\leq 8.9$ GeV$^{-2}$, wherever the interval is shorter, it means that no solution was available beyond that point. In particular, DIP$^{(1)}_{\hat\gamma;\chi_0}$ gave no solution at all in the full interval.

Taken at face value, the results of Table~IV indicate a potential $(10-15)\%$ effect due to the our present knowledge of the parameter $\chi_0$. The way to read this result is that the off-shellness of the pion can induce, at most, a $15\%$ shift on values obtained within the pion-pole approximation. While the uncertainty on $\chi_0$ can only be reduced with a reliable calculation (probably from the lattice), we can still investigate which are the preferred values for $\chi_0$ for the different interpolators we have been dealing with. The results are collected in Table~V. Notice that lower values for $\chi_0$ are preferred (first six rows). In contrast, large values for $\chi_0$ give wrong predictions for the low-energy parameters (see the last 3 rows) and hence they are clearly disfavored. While this might be indicative, it is certainly not conclusive. It is therefore essential to have a better understanding of $\chi_0$. Incidentally, Table~V also shows that DIP$_{\hat{\gamma}}^{(1,2)}$ predicts not only slightly large values for $\hat\beta$ but also negative central values for $\chi_0$. This seems to confirm that the results of DIP$_{\hat{\gamma}}^{(1,2)}$ might be unreliable.


\end{document}